\documentclass[%
 aip,
 superscriptaddress,
 amsmath,amssymb,
 reprint,%
]{revtex4-2}

\usepackage{graphicx}
\usepackage{bm}

\usepackage[utf8]{inputenc}
\usepackage[T1]{fontenc}
\usepackage{mathptmx}
\usepackage{etoolbox}
\usepackage{float}
 \usepackage{xspace}

\makeatletter
\def\@email#1#2{%
 \endgroup
 \patchcmd{\titleblock@produce}
  {\frontmatter@RRAPformat}
  {\frontmatter@RRAPformat{\produce@RRAP{*#1\href{mailto:#2}{#2}}}\frontmatter@RRAPformat}
  {}{}
}%
\makeatother
\newcommand{\betan}{\ensuremath{\beta_\mathrm{N}}\xspace}

\begin{document}

\preprint{AIP/123-QED}

\title[The Negative Triangularity Tokamak Path]{The Negative Triangularity Tokamak Path for Fusion Pilot Plants: Experimental Progress and Future Prospects}

\author{K. E. Thome}
 \email{thomek@fusion.gat.com}
\affiliation{General Atomics, San Diego, USA}

\author{M. E. Austin}
\affiliation{Institute for Fusion Studies, University of Texas – Austin, Austin, USA}

\author{S. Coda}
\affiliation{Ecole Polytechnique Fédérale de Lausanne (EPFL), \\ Swiss Plasma Center (SPC), Lausanne, CH}

\author{A. Marinoni}
\affiliation{University of California, San Diego, San Diego, USA} 

\author{A. Hyatt}
\affiliation{General Atomics, San Diego, USA}

\author{A. O. Nelson}
\affiliation{Columbia University, New York, USA}  

\author{T. Odstr\v{c}il}
\affiliation{General Atomics, San Diego, USA}

\author{C. A. Paz-Soldan}
\affiliation{Columbia University, New York, USA}  

\author{O. Sauter}
\affiliation{Ecole Polytechnique Fédérale de Lausanne (EPFL), \\ Swiss Plasma Center (SPC), Lausanne, CH}

\author{F. Scotti}
\affiliation{Lawrence Livermore National Laboratory, Livermore, USA}

\author{B. Vanovac}
\affiliation{Plasma Science and Fusion Center, \\ MIT, Cambridge, USA}
\begin{abstract}
This paper reviews the experimental progress of negative triangularity (NT),  a tokamak configuration where the poloidal cross-section is a reversed-D shape compared to the conventional positive triangularity (PT) shape. NT is a promising reactor scenario that addresses the fundamental tension between performance, exhaust, and robustness. NT studies have accelerated globally across these three pillars over the past several years. While tokamak pilot plants are typically designed for the standard PT H-mode regime, this approach faces significant challenges in balancing high core performance with manageable heat and particle exhaust as well as reliable robustness. In contrast, NT plasmas have achieved H-mode-level confinement while remaining robustly free of the deleterious edge localized mode (ELM) instability. Regarding exhaust, NT offers a larger divertor wetted area on the outboard side and demonstrates compatibility with detachment and operation at high core radiation fraction  without the constraints of the L-H power threshold, while also exhibiting low core impurity retention. NT operates with high reproducibility over a wide operating space, demonstrated by robust discharge-to-discharge consistency, and has access to plasmas with very high Greenwald fractions and/or low edge safety factors compared to PT H-mode plasmas. Further research is required to answer outstanding questions related to reactor confinement extrapolation, the optimal triangularity for a reactor, and core-edge integration. NT studies in existing and planned tokamaks are increasing, as is interest in possible reactor concepts. The unique physics and engineering advantages of NT offer a robust and simplified foundation for a viable fusion power plant. 
\end{abstract}

\maketitle

\section{\label{sec:intro}Introduction}

Tokamaks are currently the leading magnetic confinement concept for fusion energy due to their superior confinement and demonstrated plasma performance. As shown in the schematic of a typical tokamak (Fig. 1), the toroidal field coils provide the strong confining field in the toroidal direction, which follows the long way around the torus. Meanwhile, the plasma current and poloidal coils generate the poloidal field in the short way around the plasma cross-section. The superposition of these magnetic fields creates the helical magnetic twist that the particles follow; an example of a segment of one particle orbit is illustrated in Fig. 1.  In this figure, poloidal coils are located on both the inboard and outboard sides of the plasma, but some tokamaks feature poloidal coils exclusively on the outboard side. Generally, increasing the number of poloidal coils allows more flexibility in plasma shaping.

Early tokamaks all had nearly circular poloidal cross-sections, as illustrated by Fig. 2(a). Non-circular cross-sections began to be explored  both experimentally and theoretically in the 1960s through the 1980s [see Ref.\cite{MarinoniRMPP} and references therein]. There are two common shaping parameters used in tokamaks: elongation ($\kappa$) and triangularity ($\delta$). An example of an elongated plasma is shown in Fig. 2(b), where the elongation is defined as the ratio of the semi-vertical axis $b$ to the minor radius $a$: $\kappa$=$b/a$. In Fig. 2, $\kappa$=1.7 in Fig. 2(b--d) and $\kappa$=1 in Fig. 2(a).
\begin{figure}[H]
\includegraphics[width=1\linewidth, trim=17cm 8.5cm 11.2cm 7cm,
    clip] {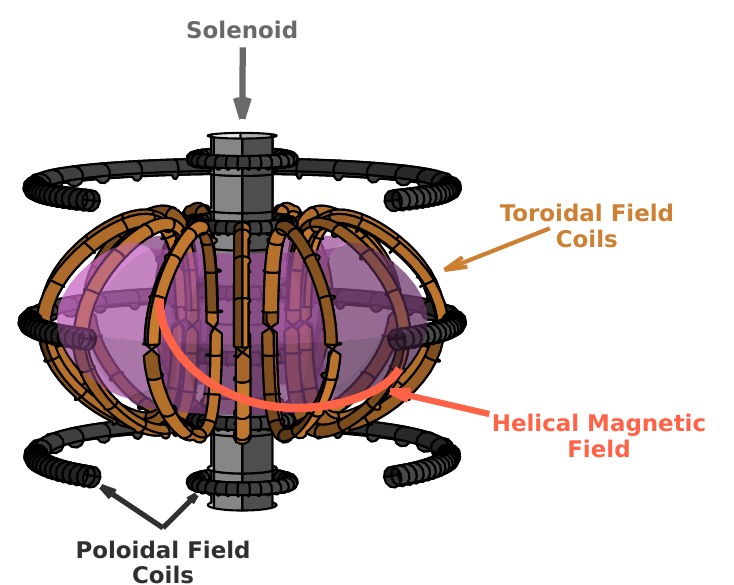}
\caption{\label{fig:figure1} Tokamak schematic with toroidal field coils (gold), poloidal field coils on both the inboard and outboard side (dark grey), and the central solenoid (grey). The  plasma is shown in maroon with the vacuum vessel omitted for clarity. The toroidal direction follows the major circumference of the tokamak and the poloidal direction follows the minor circumference encircling the plasma. }
\end{figure} 
Since tokamak plasmas are often vertically asymmetric, their cross-sections are typically characterized by distinct upper and lower triangularities ($\delta_{u/l}$), defined as $\delta_{u/l} = (R_{geo} - R_{Zmax/Zmin})/a$, where $R_{geo}$, $R_{Zmax}$ and $R_{Zmin}$ are the major radii of the geometric center and the vertical extremes, respectively. The overall average triangularity of the plasma is then defined as their mean, $\delta=(\delta_u +\delta_l)/2$. Unless stated otherwise, the triangularity referred to throughout this paper will be the average triangularity. The plasmas in Fig. 2(c) and 2(d) are vertically symmetric, with $\delta=0.4$ in Fig. 2(c) and $\delta=-0.4$ in Fig. 2(d). 

By the 1980s, advancements in magnetohydrodynamic (MHD) modeling enabled researchers to evaluate the impact of plasma shape on  stability, demonstrating that non-circular shapes provide higher stability limits [see Ref.\cite{MarinoniRMPP} and references therein]. Separately, the Dee-shape (increased triangularity) magnetic coil was found to be optimal for minimizing stresses in the plane of the coil \cite{princetond}.  The poloidal cross-sectional Dee-shape provided in Fig. 2(c) is commonly referred to as positive triangularity (PT) and is currently the shape planned for most tokamak fusion pilot plant (FPP) designs. Conversely, the shape shown in Fig. 2(d) is known as negative triangularity (NT), historically also referred to as reversed or inverse triangularity. 

\begin{figure}
\includegraphics[width=0.8\linewidth, trim=0cm 0cm 0cm 0cm,
    clip] {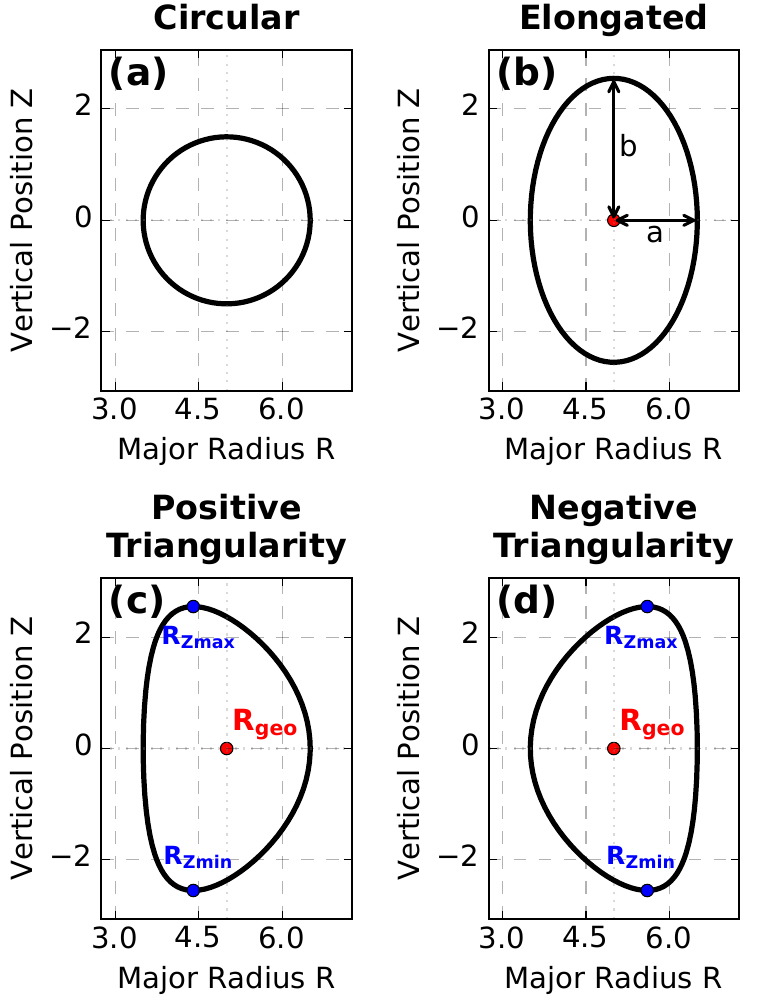}
\caption{\label{fig:figure1} Poloidal tokamak cross sections: (a) circular, (b) elongated, (c) positive triangularity and (d) negative triangularity. Shaping parameters minor radius ($a$), semi-vertical axis ($b$), geometric major radius ($R_{geo}$), and major radii of the vertical extremes ($R_{Zmax}$ and $R_{Zmin}$) are noted.} 
\end{figure}

NT was originally studied both experimentally and theoretically in the 1970s and 1980s and was then set aside due to the perceived benefits of PT at the time, which will be described in detail in the next section. Recent tokamak experiments worldwide have revived NT as a viable reactor concept. These results demonstrate a promising tokamak FPP scenario with high performance while passively alleviating long-standing power exhaust challenges, as will be described in this review paper. While a recent paper reviewed the history of NT both experimentally and theoretically up to 2021 \cite{MarinoniRMPP}, significant experimental progress has been made in the intervening years, on which this paper will focus. The remainder of this review paper is organized as follows:  Section II provides further background on tokamak physics. Section III, the bulk of this paper, details the experimental progress of NT from the 1980s to the present. Section IV discusses future work and the path to an FPP, and Section V provides concluding remarks.  

\section{\label{sec:back}Background on Tokamaks}

Achieving high plasma performance is critical to meeting the Lawson criterion \cite{lawson1955some} and reducing the cost of a reactor \cite{Wade17022021}. Considerable progress has been made towards these goals since the beginning of fusion research \cite{wurzel}. However, while accessing high plasma performance is essential, a viable reactor must also simultaneously achieve two other critical objectives: particle and heat exhaust fluxes compatible with material limits (exhaust) and a reproducible, reliable, and easily accessible plasma scenario sustained for a sufficiently long time (robustness). Thus, a sustainable fusion reactor requires this triad of performance, exhaust, and robustness, as illustrated in Fig. 3. Generally throughout this paper, performance refers to plasma performance unless otherwise specified. Tokamaks have numerous operational scenarios with many different pressure and current profiles that tend to optimize one pillar at the expense of the others. Although these existing tokamak scenarios have made substantial progress toward these goals, a fully reactor-compatible regime has not yet been achieved. As will be discussed throughout this paper, NT has recently demonstrated major advancements across performance, exhaust, and robustness. We will now discuss each of these pillars in greater detail and then summarize the current status of the conventional PT approach. 

\begin{figure}
\includegraphics[width=.7\linewidth, trim=0cm 0cm 0.8cm 0cm,
    clip] {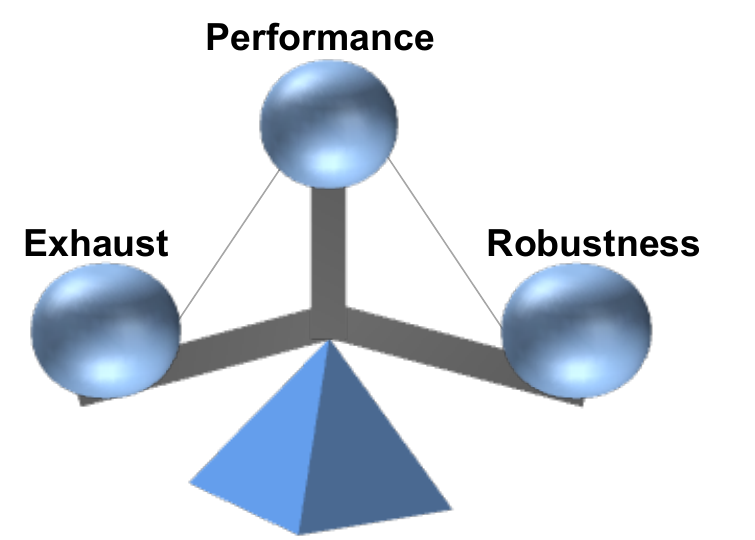}
\caption{\label{fig:figure1} Sustainable fusion reactor scenario triad of performance, exhaust and robustness.}
\end{figure}

\subsection{\label{sec:perform}Performance}
In a tokamak, there are two general types of plasma performance regimes. Low confinement (L-mode) is the baseline, starting operating regime with linear profiles, as shown in Fig. 4, and high levels of microturbulence \cite{ITERTransport_1999}. In PT, the high confinement (H-mode) regime occurs when the applied heating power exceeds the power threshold \cite{Wagner_2007}. A comparison of the edge radial electron profiles between L-mode and H-mode is shown in Fig. 4; in H-mode, an edge transport barrier or "pedestal" forms that allows for significant improvement in particle and energy confinement compared to L-mode, usually by a factor of 2. There are many different types of H-mode scenarios that have been investigated on tokamaks worldwide. Since H-mode's discovery in the ASDEX tokamak in 1982 \cite{Wagner1982}, it has been the focus of tokamak research and it is currently the planned regime for most tokamak FPPs. 

There are two widely used performance metrics in tokamaks: 1) normalized beta $\betan=\beta/(I_\mathrm{p}/aB_\mathrm{T})$, where $\beta$ is the toroidal plasma beta, $I_p$ is the plasma current, and $B_\mathrm{T}$ is the toroidal magnetic field,  and 2) the standard normalized H-mode energy confinement factor $H_\mathrm{98y,2}$ derived from multi-machine scalings \cite{ITERTransport_1999}. Target values of $\betan$ for a reactor typically vary from 1--3+\cite{Creely_2020_SPARC, SICCINIO2022113047, 10.1088/978-0-7503-2719-0ch8, Buttery_2021}, with a target value of 1.8 on both ARC \cite{Hillesheim_2026_ARC} and ITER \cite{Shimada_2007}. Plasmas with $H_\mathrm{98y,2}$ of 1 have energy confinement that is consistent with standard H-mode plasmas. $H_\mathrm{98y,2}$ of at least 1 is targeted in FPP designs with higher $H_\mathrm{98y,2}$ projected to reduce costs \cite{Wade17022021}.  Note, other operational and performance metrics exist, such as plasma density, edge safety factor, or the normalized $I_\mathrm{p}/aB_\mathrm{T}$, among others. 

\begin{figure}
\includegraphics[width=0.7\linewidth, trim=0cm 0cm 0cm 0cm,
    clip] {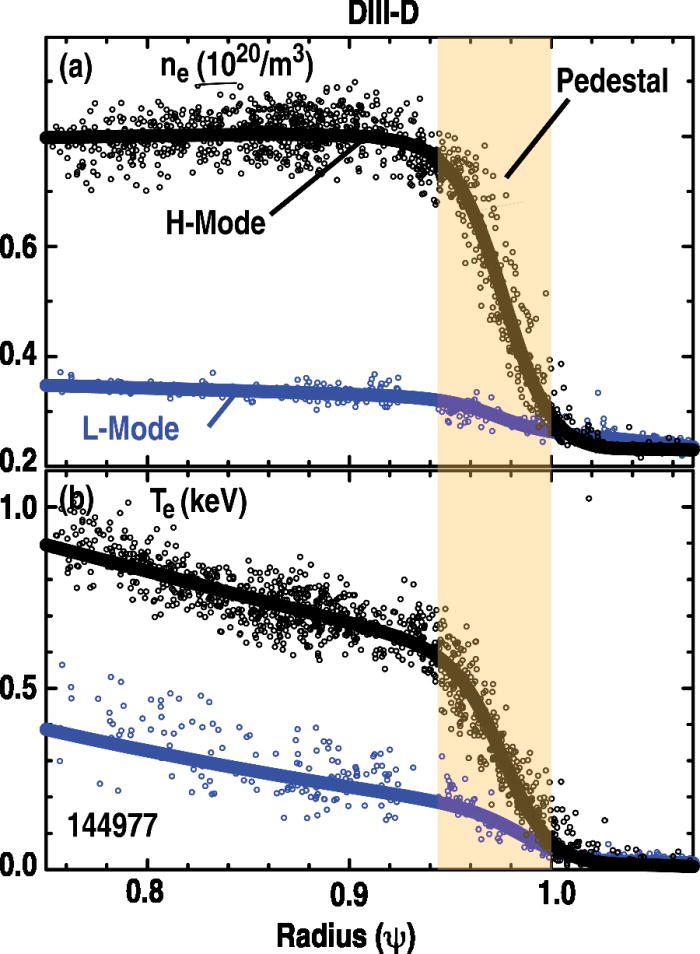}
\caption{\label{fig:figure3} Reproduced from \cite{Leonard_ELMreview}. Edge radial electron profiles measured by Thomson scattering on \mbox{DIII-D} of (a) electron density and (b) electron temperature of L-mode (blue) and H-mode (black) plasmas with a pedestal observed in the H-mode plasma. }
\end{figure}

\begin{figure*}[t]
\includegraphics[width=1\linewidth, trim=0cm 0cm 0cm 0cm,
    clip] {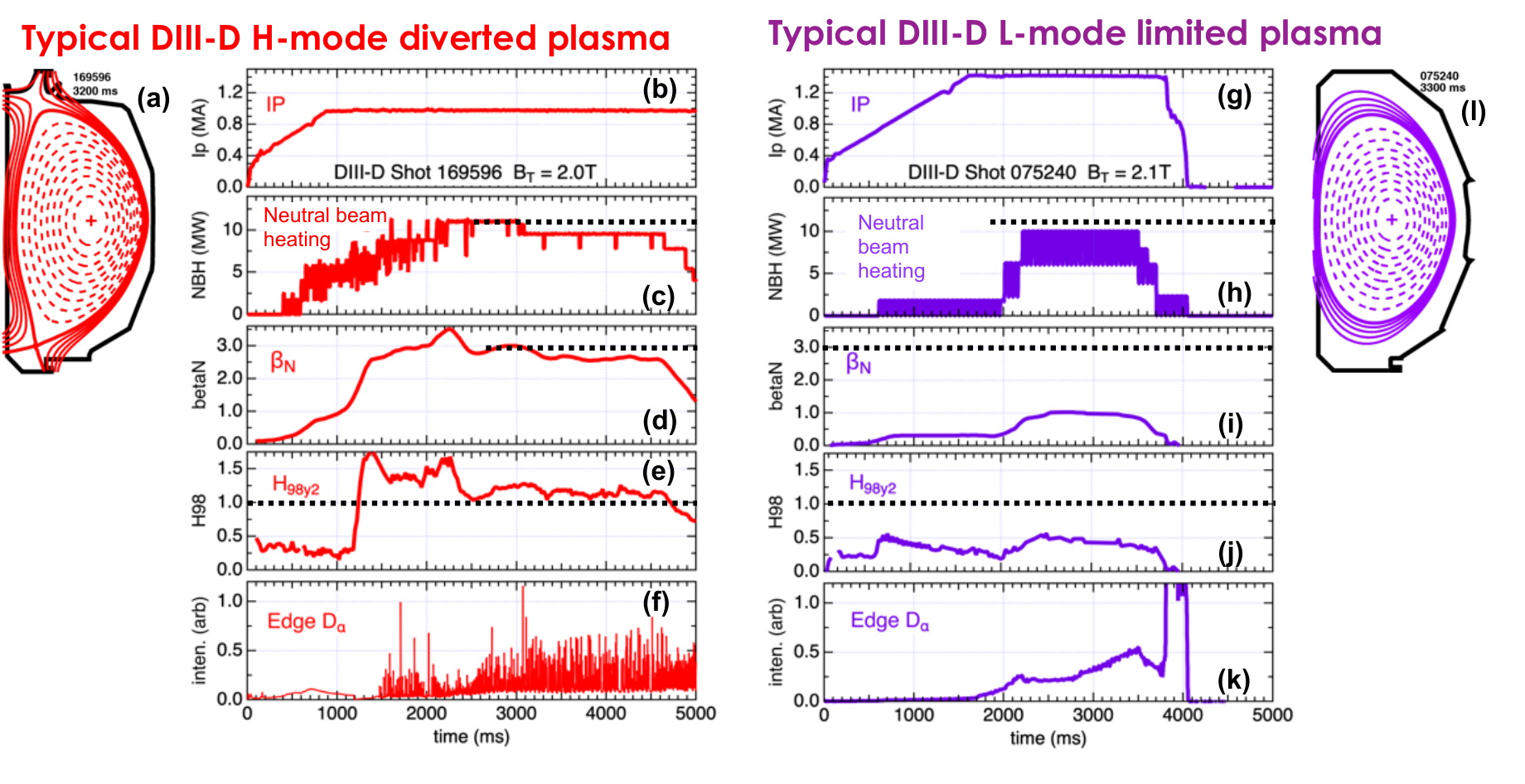}
\caption{\label{fig:figure3} Courtesy of M. Austin. Comparison of a \mbox{DIII-D} H-mode diverted plasma (left) with a L-mode limited plasma (right). The equilibrium shapes are given in (a) and (l), plasma current in (b) and (g), neutral beam heating in (c) and (h),  $\betan$ in (d) and (i), $H_\mathrm{98y,2}$ in (e) and (j) and edge $D_\alpha$ in (f) and (k) for the H-mode (red) and L-mode (purple) discharges, respectively. }
\end{figure*}
A comparison of two \mbox{DIII-D} discharges is given in Fig. 5. These plasmas are in two different magnetic configurations: the plasma in Fig. 5(a) is in the diverted configuration and the plasma in Fig. 5(l) in the limited configuration. In the limited topology, the last closed flux surface (LCFS) is defined solely by a solid surface (i.e. limiter), whereas, in the diverted case, the LCFS is defined solely by the magnetic field and requires extra poloidal coils to make that shape. There are many different types of diverted topologies; in this paper the focus will be on plasmas with only one null or X-point. Fig. 5(a) is an example of a lower single-null (LSN) diverted plasma.  These two discharges have comparable plasma current [Fig. 5(b) and Fig. 5(g)] and neutral beam heating [Fig. 5(c) and Fig. 5 (h)] but the diverted configuration allows access to H-mode. The L-H power threshold is known to be generally lower in diverted plasmas compared to limited plasmas \cite{burrell1989confinement, Carlstromlh}. The two performance metrics are also shown for both plasmas; in the H-mode plasma $\betan$ is slightly lower than 3 [Fig. 5(d)] whereas in the L-mode plasma $\betan$ is around 1. The H-mode plasma achieves $H_\mathrm{98y,2} >1$, whereas the L-mode plasma remains near 0.5. Thus, relevant $\betan$ and $H_\mathrm{98y,2}$ are only obtained in the H-mode case in PT. 

\subsection{\label{sec:exhaust}Exhaust}

Controlling the heat and particle exhaust is a central challenge in an FPP. Diverted tokamak configurations are the planned topology for a reactor. The power leaving the plasma core in current tokamaks is 10--20 MW, but in a tokamak reactor it will be greater than 100 MW. This power flows through a narrow 1--10 mm scrape-off layer toward the X-point and divertor targets as shown by the ITER schematic in Fig. 6. Regarding the heat exhaust, plasma facing component materials can only tolerate 5-10 $\mathrm{MW/m^2}$ before they melt, so this power must be dissipated before reaching material surfaces via radiation to reduce the heat flux and minimize sputtering. Radiative divertor conditions must be maintained with minimal core performance degradation. Particles must also be exhausted. In a reactor, this includes unused tritium fuel alongside several types of impurities: helium ash from the fusion reactions and both intrinsic and extrinsic impurities. Low-Z impurities dilute the main fuel, thereby reducing fusion reactions, whereas high-Z impurities can cause significant radiation losses. Finally, both steady-state and transient heat and particle fluxes can occur in a tokamak and must be mitigated in a reactor. 

\begin{figure}
\includegraphics[width=1\linewidth, trim=0cm 0cm 0cm 0cm,
    clip] {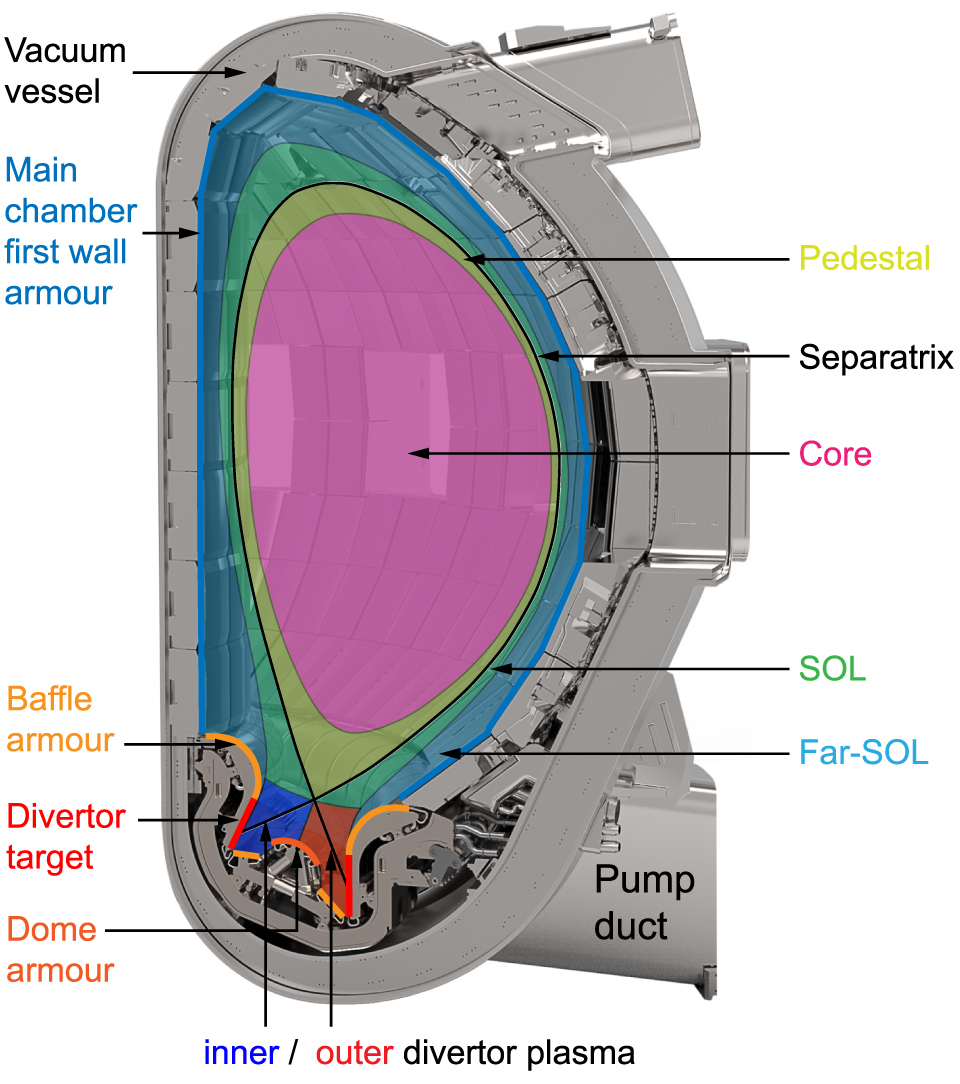}
\caption{\label{fig:figure3} Reproduced from \cite{Krieger_2025}. CAD view of ITER vessel sector illustrating the main plasma facing components and principal plasma regions. }
\end{figure}
\subsection{\label{sec:repro}Robustness}

A high capacity factor (the ratio of actual energy output to maximum possible output) is important for economic viability. Commercial fission currently has a capacity factor of over 90\% in the United States\cite{eia_epm_2025}; this value has increased with time from approximately 60\% in the late 1970s and 1980s \cite{ans_capacity_2019} as operators gained experience. Maintenance and unexpected shutdowns decrease the capacity factor and increase the cost and complexity of the reactor. Reproducible, reliable plasmas are essential for power generation at lower costs, and plasma performance should not come at the expense of uptime. 

While achieving these commercial-grade capacity factor metrics is not an immediate requirement for present-day devices, they serve as critical long-term goal posts that must inform current scenario selection. A primary hurdle in modern magnetic fusion research is that the community is currently operating and optimizing within relatively small-scale, short-pulse experimental facilities, rather than true power-producing reactors. Since the underlying physics can change significantly when transitioning to larger device dimensions, the path toward scaling these localized plasma configurations to power-plant levels remains a critical uncertainty. While sophisticated plasma and engineering solutions to current reproducibility challenges in some scenarios offer great promise for stabilizing localized operating windows, relying on these active techniques introduces scaling uncertainties. 

To navigate these scaling challenges, the fusion community must address a phenomenon common to many frontier experiments: the highest-performance discharges in fusion, which are widely reported, are frequently transient "trophy" discharges that can be challenging to reliably replicate. This phenomenon is not unique to any single facility or specific confinement concept; rather, it is a universal characteristic across experimental physics and broader fusion research, where pushing the absolute frontier of a device's capability naturally introduces high operational sensitivity. In tokamaks and other magnetic fusion devices, these trophy discharges often have marginal stability against MHD or edge instabilities, making sustained operation at high performance unreliable. 

Furthermore, these scenarios are currently often only accessible within a very narrow operational space that can be heavily constrained by other parameters such as exact shaping, wall and vacuum conditions, and other sensitive variables that can be difficult to quantify. Consequently, even minor changes in these baseline conditions can cause plasma performance to degrade or terminate prematurely. For a fusion reactor, the chosen operational regime must be easily accessible, feature a wide operating space, and be reliably reproducible.

\begin{figure}
\includegraphics[width=0.9\linewidth, trim=0cm 0cm 0cm 0cm,
    clip] {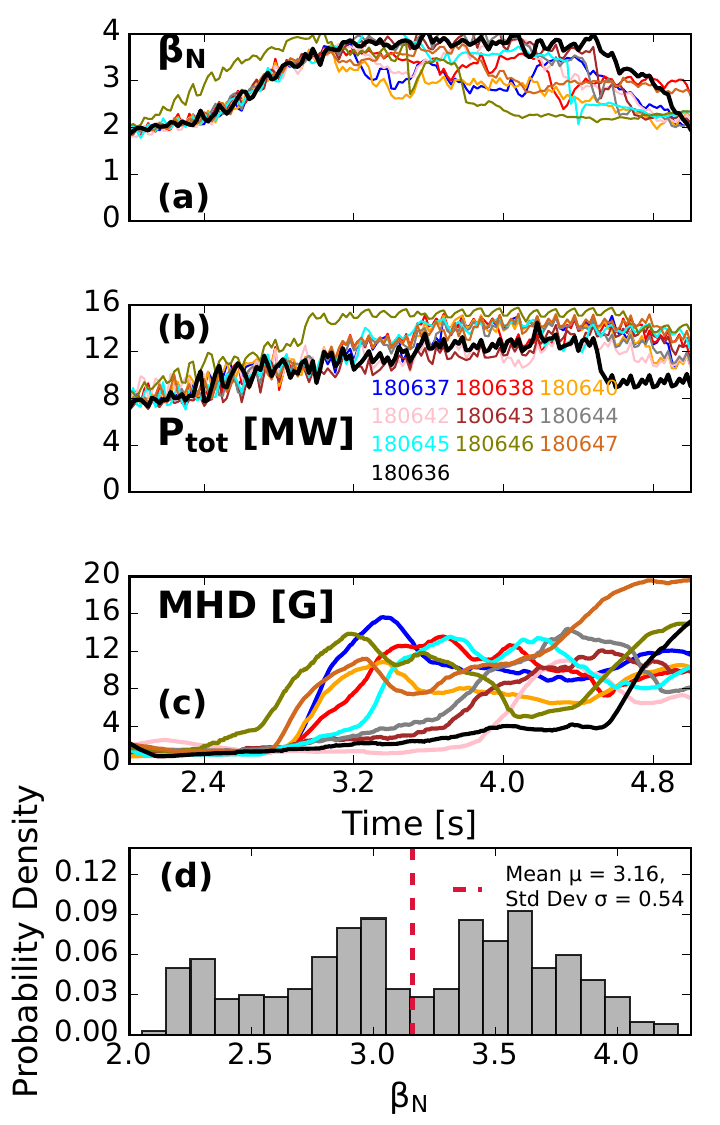}
\caption{\label{fig:figure7} Reproducibility of a very high-performance PT H-mode \mbox{DIII-D} discharge (180636 in thick black) compared to other discharges during that half run day with equivalent applied power: (a) $\betan$, (b) total auxiliary applied power $P_\mathrm{tot}$, (c) $n=1$ MHD fluctuation amplitude, and (d) the PDF of the operating distribution from these plasmas, where $P_\mathrm{tot}\geq$12 MW.   }
\end{figure}

As an example of this "trophy" discharge phenomenon, a \mbox{DIII-D} PT H-mode discharge that achieved and sustained a very high $\betan$ of approximately 4 is shown in Fig. 7(a) by the thick black line. This specific experimental half run day was designed to explore the performance limits of this scenario by applying high levels of auxiliary power to build upon a previous discharge. Alongside the trophy discharge are the other discharges from that half run day that had equivalent or higher applied power [Fig. 7(b)]. As can be seen in Fig. 7(a), no other discharge repeated the high $\betan$ trajectory of the "trophy" discharge, but instead many of them were limited by MHD instabilities, with $n=1$ being the most limiting as shown in Fig. 7(c). However, all discharges sustained a $\betan$ of 2 or higher during the flattop period and 58\% of the time they sustained high $\betan$ of at least 3. This illustrates that while moderate-to-high performance is readily reproducible, the highest performance is not. The operational distribution of this scenario is captured statistically by the probability density function (PDF) shown in Fig. 7(d), which represents the total accumulated time across all ten discharges where a high-power flat-top was successfully maintained ($P_{\text{tot}} > 12$\,MW) for each given value of $\beta_N$. The resulting mean of 3.16 and wide standard deviation of $\sigma = 0.54$ reflect a plasma state that naturally spans a broad envelope of performance levels as it navigates the marginal stability boundaries inherent to this operating regime under high auxiliary heating.

\subsection{\label{sec:repro}Current Status of Sustainable Tokamak Reactor Scenarios}

H-mode is currently the planned tokamak reactor operating regime since it satisfies the performance pillar (high confinement) required for a fusion reactor. Ongoing research seeks to further increase performance in H-mode and thus lower the cost. However, H-mode exacerbates the heat and particle exhaust challenges in three key ways. First, the power crossing the separatrix must be higher than the H-mode power threshold. This threshold scales with density, toroidal field, and plasma boundary surface area\cite{Martin_2008}, which are all expected to increase in a reactor compared to existing machines, significantly raising the baseline power that must then be exhausted. Second, H-mode introduces the edge localized mode (ELM) instability \cite{Leonard_ELMreview} that transiently expels heat and particles, as is represented by each of the spikes in the edge $D_\alpha$ filterscope in Fig. 5(f). While ELMs are generally inconsequential in today's experiments, the projected ELM damage scales severely with reactor size. Thus, an ELM-free or mitigated ELM regime is required for a reactor\cite{EVANS2013S11,Paz-Soldan_2021,VIEZZER2023101308}. However, achieving these states usually entails a confinement penalty; the active control techniques or inherent edge fluctuation modes required to suppress ELMs typically induce additional transport that degrades overall performance. Third, due to the formation of the edge particle transport barrier in H-mode, the impurity confinement time is too long, causing impurities to accumulate in the plasma core and affect the performance, as will be described in Sec. IIIB.

Finally, regarding the third and often completely ignored pillar, high-performing H-mode plasmas can be challenging to sustain, reproduce, and reliably access. Faced with these profound physics and engineering trade-offs --- where the performance pillar is heavily favored at the direct expense of exhaust and robustness --- a critical question arises: is standard PT H-mode the only way forward? The rest of this paper will discuss the benefits of NT as a sustainable reactor scenario. A comparison of PT H-mode and NT is given in Table 1; as can be seen from that table, NT achieves sufficient confinement, sidesteps the L-H power threshold, is robustly ELM-free, has sufficient impurity confinement, and exhibits good robustness across plasma discharges. NT also has a more accessible divertor and large exhaust area.  

\begin{table}[tbp]
\centering
\caption{\label{tab:comparison_table} Comparison of the tokamak H-mode and NT regimes against the sustainable reactor triad.}
\renewcommand{\arraystretch}{1.4} 
\begin{tabular}{| p{0.31\linewidth} | p{0.37\linewidth} | p{0.26\linewidth} |}
\hline
 & \textbf{H-mode} & \textbf{NT} \\
\hline
\textbf{Performance} & {\raggedright Highest confinement\par} & {\raggedright Sufficient confinement\par} \\
\hline
\textbf{Exhaust:} & & \\
\hline
{\raggedright \hspace{0.2cm}\hangindent=0.2cm\hangafter=1 L-H Power Threshold\par} & Yes & No \\
\hline
{\raggedright \hspace{0.2cm}\hangindent=0.2cm\hangafter=1 Robustly ELM-free\par} & No & Yes \\
\hline
{\raggedright \hspace{0.2cm}\hangindent=0.2cm\hangafter=1 Impurity Confinement\par} & {\raggedright Too good in most scenarios\par} & {\raggedright Sufficient\par} \\
\hline
{\raggedright \textbf{Robust Plasma} \newline \textbf{Scenario}\par} & {\raggedright Reproducible at intermediate performance, poor at highest performance; operational space is highly regime-dependent\par} & {\raggedright Reproducible plasmas over a wide operational space\par} \\
\hline
\end{tabular}
\end{table}

\section{\label{sec:progress}Negative Triangularity Experimental Progress}

PT ($\delta>0$) is the conventional tokamak poloidal cross-section shape, as shown in the bottom of Fig. 8. NT ($\delta<0$) is a plasma shape where the uppermost and lowermost points of the cross-section are shifted toward the outboard side, creating a reversed-D shape, as shown in the top of Fig. 8.  As illustrated, the inboard side of the torus represents the favorable magnetic curvature region, while the outboard side represents the unfavorable curvature region, where the magnetic field curvature and pressure gradient align to destabilize curvature-driven instabilities, ranging from macro-scale ballooning modes to micro-scale drift-interchange modes. Geometrically, NT places a larger fraction of the plasma volume in this unfavorable region compared to PT, as seen in Fig. 8. Counterintuitively, despite this geometric disadvantage, NT plasmas exhibit robust stability and high confinement, as will be explored throughout the remainder of this review. 
\begin{figure}
\includegraphics[width=.8\linewidth, trim=0cm 0cm 0cm 0cm,
    clip] {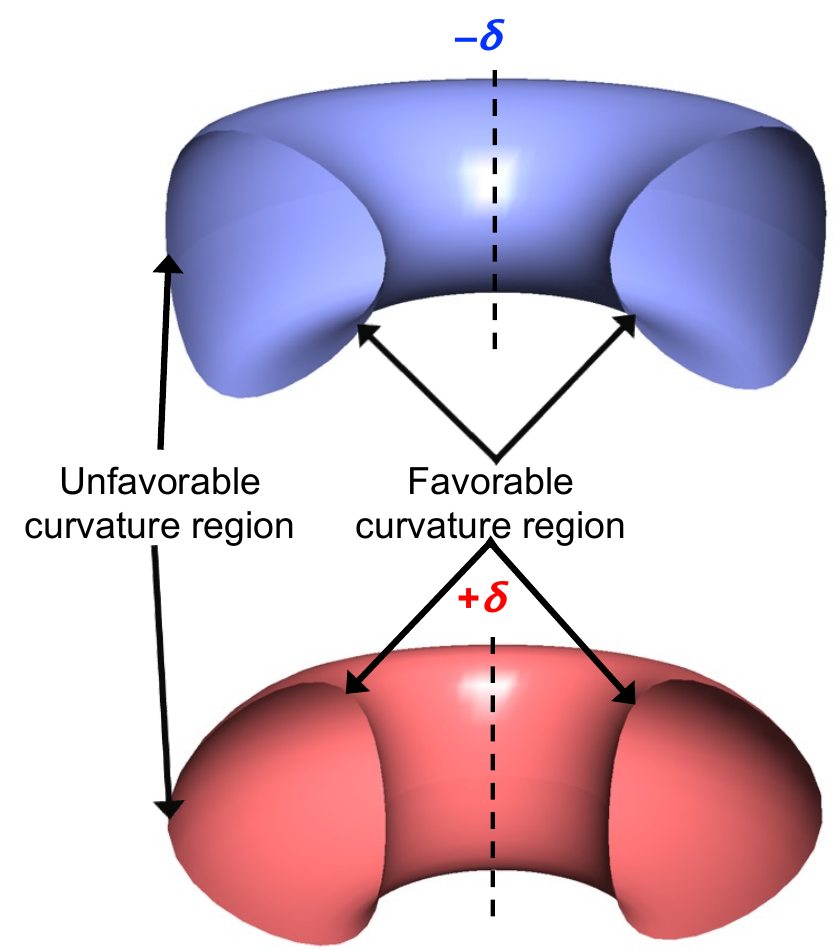}
\caption{\label{fig:figure7} Courtesy of M. Austin. 3D renderings of tokamak poloidal cross-sections for NT (top) and PT (bottom) configurations, highlighting the favorable and unfavorable magnetic curvature regions.}  
\end{figure}

This section surveys the experimental progress in NT research. The foundational experiments from  the 1980s through approximately 2021 will be described in the Foundations subsection, while the rapid progress achieved since then will be detailed in the Recent Advances subsection.

\subsection{\label{sec:found}Foundations}

The world's first experimental studies of NT were conducted in the United States during the late 1970s/early 1980s on the Poloidal Divertor eXperiment (PDX) \cite{Meade1981PDX} at the Princeton Plasma Physics Laboratory [NT cross-section shown in Fig. 9(a)] and Tokapole II \cite{Lipschultz_1980} at the University of Wisconsin–Madison [NT cross-section shown in Fig. 9(b)]. These experiments mainly compared vertical stability between NT and PT, but they also found that the NT confinement was similar to that of PT. However, it should be noted that these plasmas had weak NT, high squareness and high collisionality (unlike today's devices.) There were no further experiments immediately following these initial studies.

\begin{figure}
\includegraphics[width=1\linewidth, trim=0cm 0cm 2.5cm 0cm,
    clip] {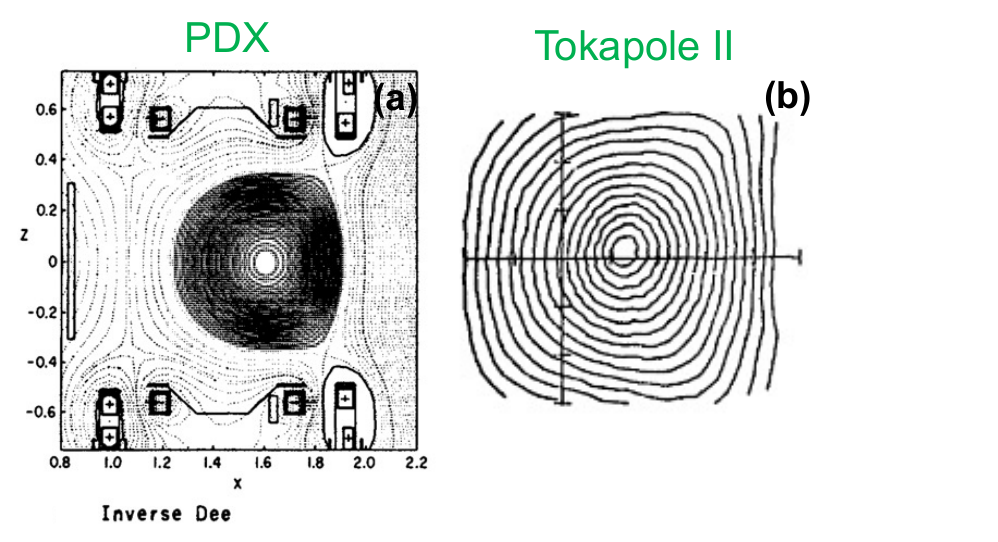}
\caption{\label{fig:figure7} Poloidal cross-sections of the world's first experimental NT plasmas in (a) PDX \cite{Meade1981PDX} and (b) Tokapole II\cite{Lipschultz_1980}.}  
\end{figure}

Since NT lacks a deep magnetic well, generally favorable curvature, it was predicted to be more prone to interchange and ballooning modes (Mercier criterion) and to have generally lower normalized beta stability limits compared to PT. Consequently, NT was abandoned experimentally in the 1980s due to this predicted lower MHD stability and the concurrent experimental discovery of H-mode. However, in the 1990s a new device came online: the Tokamak à Configuration Variable (TCV) tokamak at École Polytechnique Fédérale de Lausanne (EPFL) in Switzerland\cite{FHofmann_1994, Theiler_2026}. TCV was explicitly designed to explore strong shaping with eight poloidal field coils on both the inboard and outboard of the tokamak that are individually powered and controlled, for 16 total. 

TCV began exploring NT in the 1990s, starting originally with limiter plasmas and Ohmic heating only \cite{moret}. In 1997, at high collisionality in NT, they observed similar results to those of PDX. However, when TCV pushed to stronger NT shapes and reduced the density, they observed higher electron energy confinement ($\tau_{E_e}$) compared to PT L-modes with both Ohmic heating and electron cyclotron resonance heating (ECRH) as shown in Fig. 10(a) \cite{Pochelon1998TCV}. TCV continued their NT studies over the next decade and found that, using ECRH as their main source of auxiliary heating, confinement exhibited a strong dependence on collisionality and on $\delta$ as shown in Fig. 10(b) \cite{Camenen_2007}. The electron heat transport is distinctly lower in the strong NT (blue, $\delta$ = -0.4) compared to strong PT (red, $\delta$ = +0.4), corresponding to the improved confinement, particularly at lower collisionalities (the right part of the figure.) Note that at higher collisionalities (on the bottom left), there is little difference between PT and NT, in agreement with the initial NT results on PDX. 

\begin{figure}
\includegraphics[width=.9\linewidth, trim=0cm 0cm 1.2cm 0cm,
    clip] {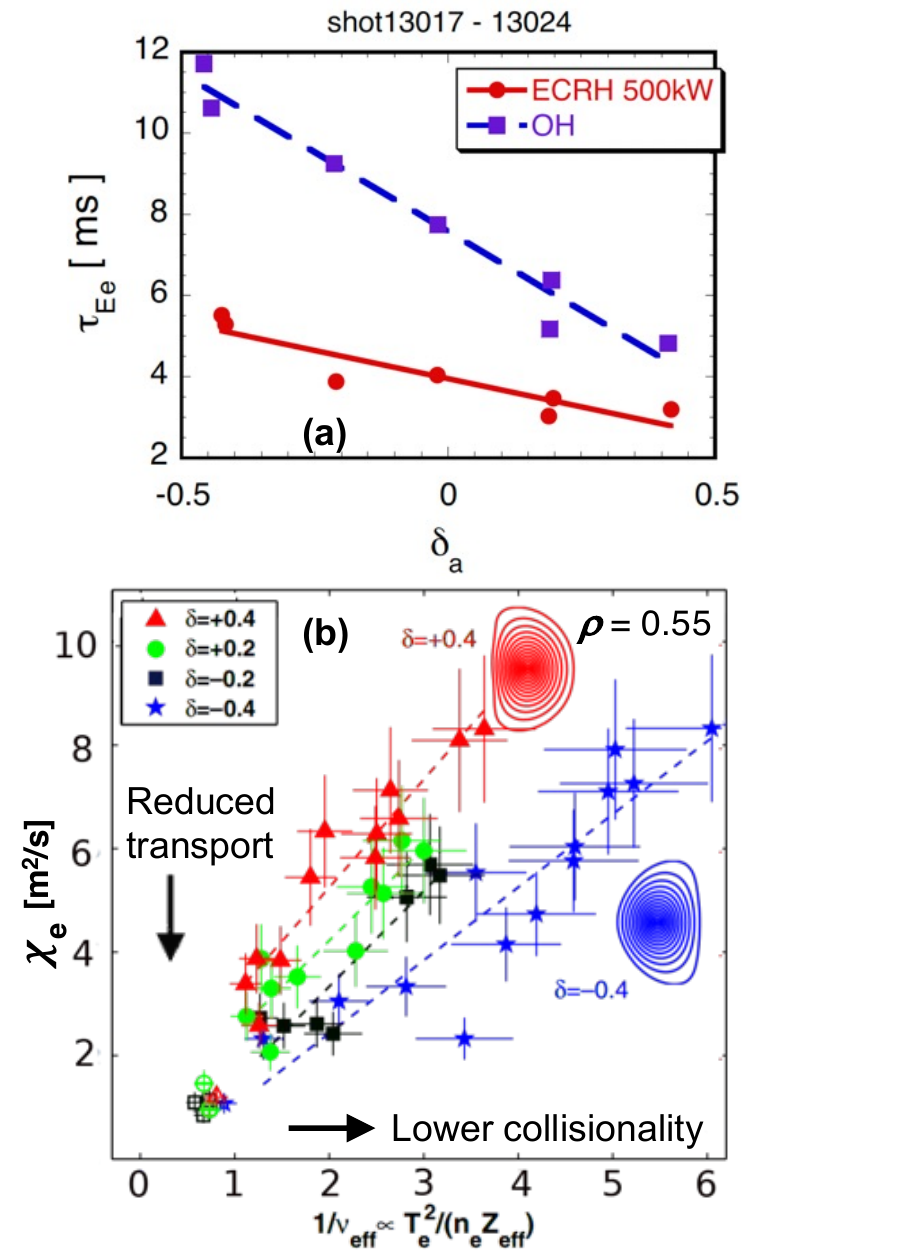}
\caption{\label{fig:figure7} Initial TCV NT Results. (a) Electron confinement time ($\tau_{E_e}$) versus triangularity ($\delta_a$) for Ohmic (OH in blue squares) and Electron Cyclotron Resonance Heating (ECRH in red circles) reproduced from \cite{Pochelon1998TCV}. Electron heat transport dependence on triangularity and collisionality adapted from \cite{Camenen_2007}. }  
\end{figure}

These initially surprising results were eventually explained by lower measured turbulence levels in NT compared to PT on TCV\cite{Huang_2019}, as shown in Fig. 11(a) for a large part of the plasma radius (from $\rho=0.5$ to 0.9). Microinstabilities are known to drive anomalous cross-field transport; thus, reduced turbulence directly correlates with better confinement. As discussed earlier, turbulence levels are generally higher in L-mode than in H-mode. Among the variety of microinstability regimes observed in tokamaks, the Ion Temperature Gradient (ITG) and Trapped Electron Mode (TEM) are prevalent electrostatic modes expected in a tokamak reactor and routinely found in today's machines. It was found that TCV NT plasmas are dominated by TEM turbulence, which is driven by a resonance between a drift wave (of either density or temperature) and the toroidal precessional drift of trapped electrons around the plasma [Fig. 11(b)] \cite{Kadomtsev_1971}. The calculated TEM growth rates are lower in NT than in PT, as shown in Fig. 11(c) \cite{Marinoni_2009}. This stabilization occurs because the trapped electrons spend more time in the favorable curvature region in NT than in PT, as illustrated in Fig. 11(d) \cite{Marinoni_2021}. Thus, the geometry of NT stabilizes the TEM turbulent mode and is responsible for improved confinement in NT. Further studies indicate that the ITG mode might also be stabilized by the NT geometry \cite{waltz, duff, merlo, Singh_2025}.

\begin{figure}
\includegraphics[width=1\linewidth, trim=0cm 0cm 0cm 0cm,
    clip] {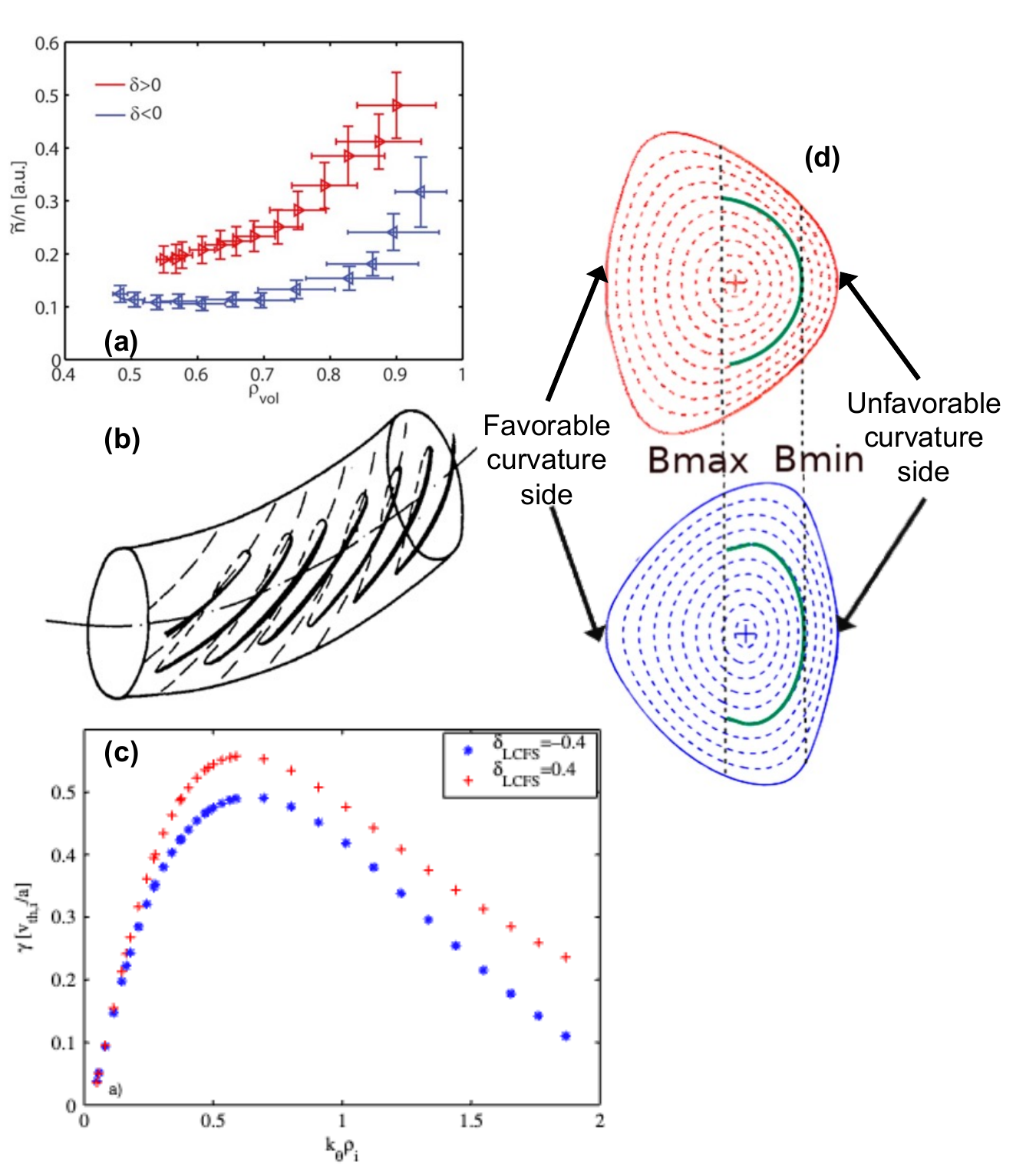}
\caption{\label{fig:figure7} TEM stabilization in NT plasmas. (a) Density fluctuations dependence on triangularity in TCV reproduced from\cite{Huang_2019}.(b) Trapped electrons following banana orbit on outside of tokamak reproduced from \cite{Kadomtsev_1971}. (c) TEM growth rates dependence on triangularity of TCV plasmas reproduced from \cite{Marinoni_2009}. (d) Trapped electron orbit in green in both  PT (top) and NT (bottom) shapes, adapted from \cite{Marinoni_2021}. }  
\end{figure}

The confinement improvements on TCV motivated further NT experiments in other tokamaks in the mid-to-late 2010s. \mbox{DIII-D}\cite{J.L.Luxon_2002} first explored NT in an inner-wall-limited shape with a strong NT of $\delta=-0.4$ [Fig. 12(a)]. Approximately 10 MW of auxiliary power from neutral beam injection (NBI) and ECRH [Fig. 12(b)] was injected into these NT plasmas. Remarkably, $\betan$ of approximately 2.5 [Fig. 12(c)] and a $H_{98y,2}$ = 1.2 [Fig. 12(d)] were simultaneously achieved in a plasma entirely without ELMs, as demonstrated by the lack of spikes in the $D_\alpha$ filterscope signal [Fig. 12(e)] \cite{Austin}. These experiments on \mbox{DIII-D} provided the first clear demonstration of sufficiently high beta in NT, directly contradicting the early pessimistic MHD stability expectations. To provide a direct comparison, a comparable PT shape was also evaluated on \mbox{DIII-D}, including L-mode cases at low power. The NT case exhibited both increased confinement and reduced lower turbulence levels than its PT counterpart\cite{Austin,Marinoni2019}. Linear gyrokinetic analysis showed that both the PT and NT plasmas are dominated by TEMs, indicating that the TEM is responsible for the improved confinement in NT on \mbox{DIII-D} \cite{Marinoni2019}.

\begin{figure}
\includegraphics[width=1\linewidth, trim=0cm 0cm 0cm 0cm,
    clip] {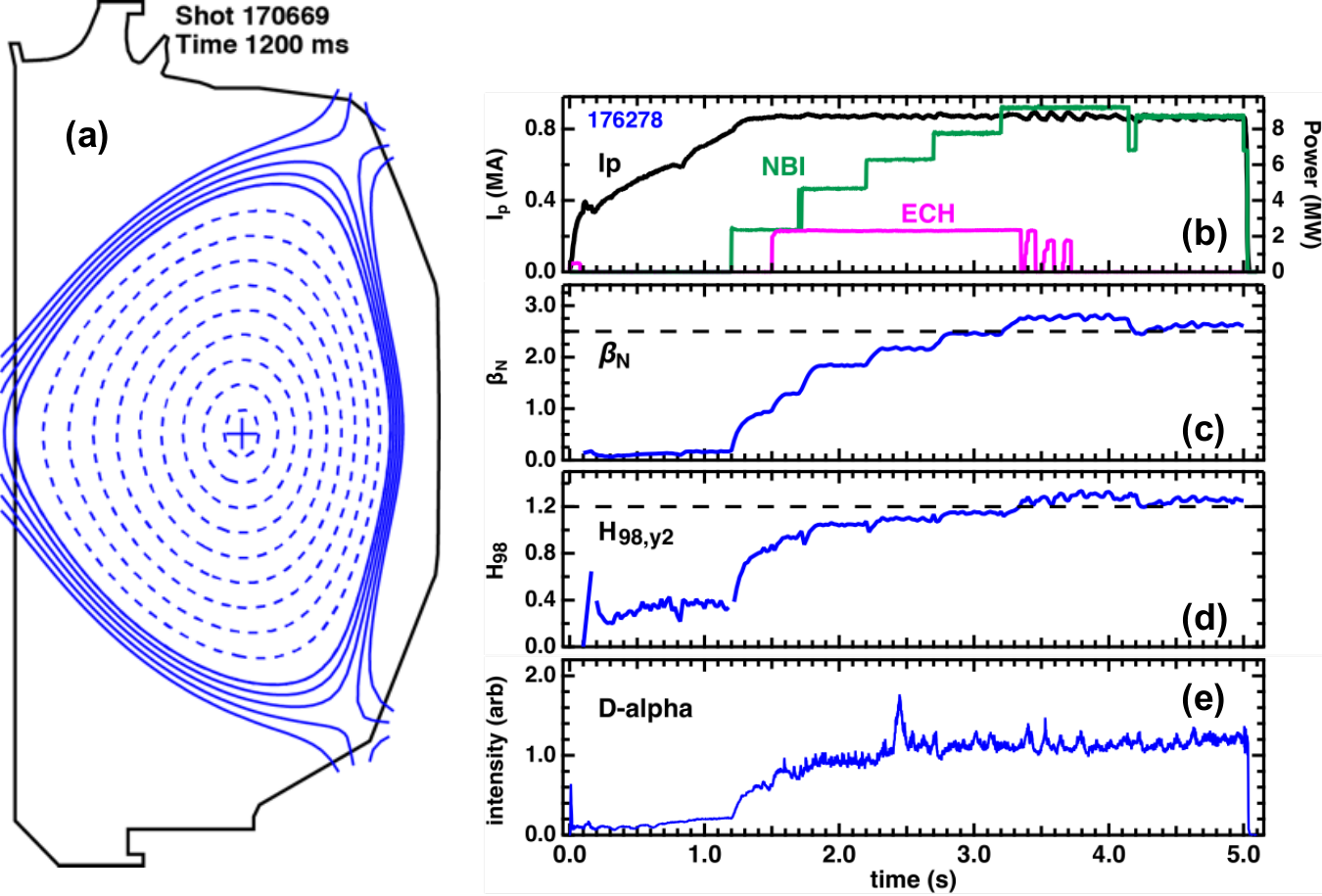}
\caption{\label{fig:figure12} Initial NT experiments on \mbox{DIII-D}. (a) Limited NT equilibrium shape. High-performance time traces in a limited \mbox{DIII-D} NT plasma without ELMs: (b) plasma current ($I_p$ in black), NBI power (green), and ECRH power (pink); (c): $\betan$, (d): $H_{98y,2}$, and (e) $D_\alpha$ filterscope. Adapted from \cite{Austin}, reproduced with permission. }  
\end{figure}

Building on these foundational successes in NT limited plasmas, the international fusion community performed more dedicated NT studies in the divertor configuration on the TCV, \mbox{DIII-D} and ASDEX-U (AUG) tokamaks. Although NT plasmas are inherently less vertically stable than PT configurations \cite{Medvedev_2015,Song_2021}, operational experience has demonstrated that robust control is achievable. Notably, dedicated studies on \mbox{DIII-D} have shown that the enhanced instability growth rates of strong NT shapes remain well within the controllable limits of the device\cite{Nelson_2023}. Also, the installation of passive conducting plates is predicted to improve vertical stability and offers a clear mitigation strategy for next-step devices\cite{Guizzo_2024}. TCV achieved record performance in NT in diverted shapes with neutral beam heating \cite{Coda_2022}. \mbox{DIII-D} extended the initial high-performance results in an ELM-free plasma in a limited shape to a weaker NT diverted shape with $\delta$=-0.2\cite{Marinoni_2021}; an example of these weakly diverted NT shapes is shown on the right side of Fig. 13. 
AUG successfully created weakly NT diverted plasmas with its tungsten wall, although its NT shaping capabilities are severely limited by the geometry of its vacuum vessel \cite{Happel_2023}. 

\begin{figure}
\includegraphics[width=1\linewidth, trim=0cm 0cm 0cm 0cm,
    clip] {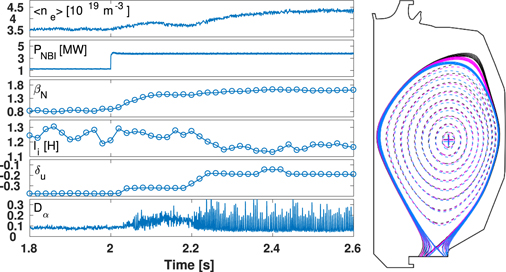}
\caption{\label{fig:figure3} Reproduced from \cite{Marinoni_2021}. Left: Time evolution of a \mbox{DIII-D} discharge that accesses H-mode. From top to bottom: line averaged density, NBI power ($P_{NBI}$), $\betan$, internal inductance ($\ell_i$), upper triangularity $\delta_u$ and $D_{\alpha}$ filterscope. Right: Three poloidal cross sections shown as the upper triangularity is relaxed from $\delta_u = -0.36$ (black) before 2.05 s, $\delta_u = -0.3$ (pink) between 2.05 and 2.2 s, and $\delta_u = -0.18$ (blue) after 2.2s }
\end{figure}

\begin{figure*}[p]
\includegraphics[width=1\linewidth, trim=0cm 0cm 2.5cm 0cm,
    clip] {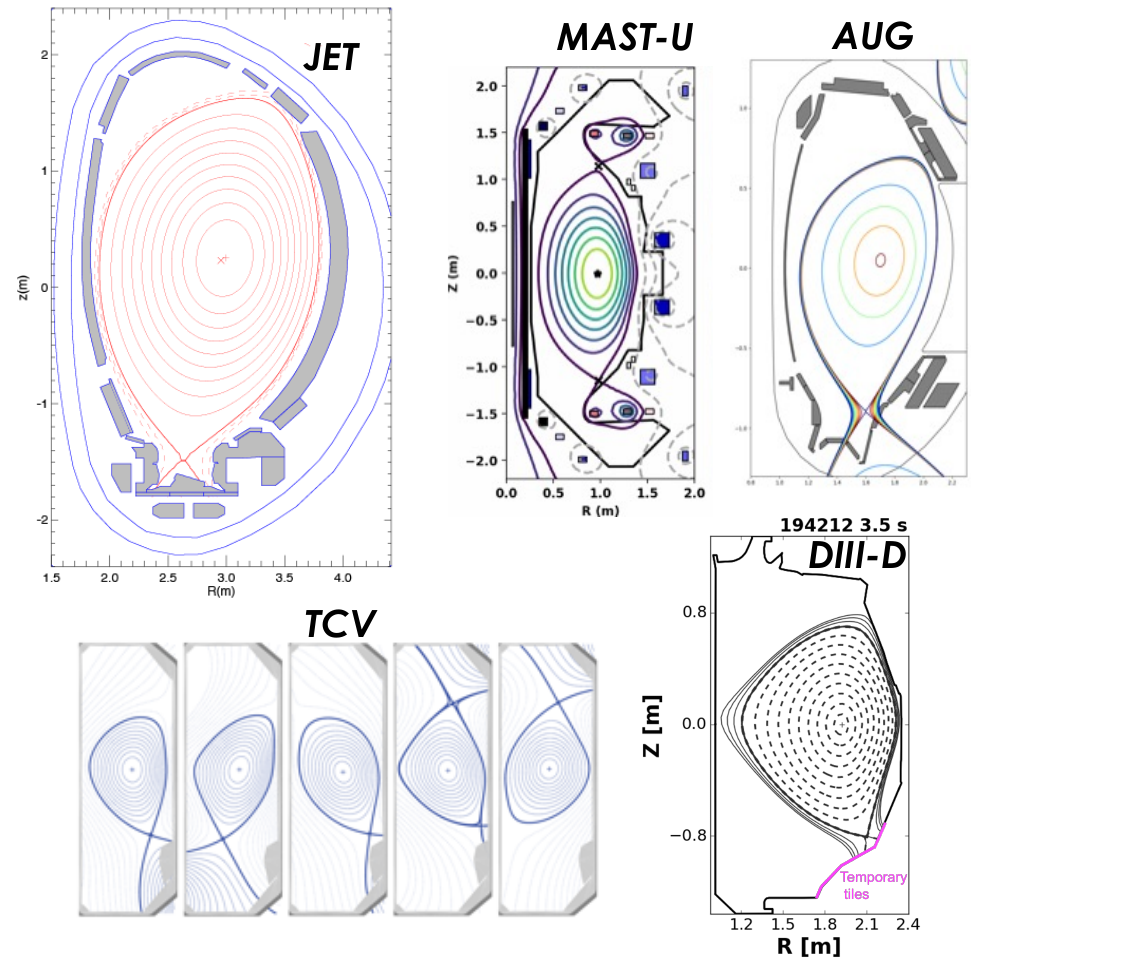}
\caption{\label{fig:figure3} Major worldwide NT experiments: JET, courtesy of O. Sauter, \mbox{MAST-U}, reproduced from\cite{Nelson_2024}, AUG, courtesy of B. Vanovac, reproduced from \cite{Vanovac_NF_submitted}, TCV reproduced from \cite{Coda_2022} and \mbox{DIII-D} adapted from \cite{Thome_2024}.}
\end{figure*}

Experiments across all three tokamaks revealed that H-mode and ELMs become accessible when the NT is sufficiently weaker than a critical value($\delta>\delta\mathrm{_{crit}}$)  \cite{Pochelon2012TCV,Marinoni_2021,Saarelma_2021, Happel_2023}. The transition to H-mode depends on various shaping parameters, with a $\delta\mathrm{_{crit}}$ existing for each device and varying somewhat between machines\cite{Nelson_2022}. An example of this L-H transition boundary during an upper triangularity scan on \mbox{DIII-D} is shown in Fig. 13. In this diverted NT shape, $\delta_l$  was held fixed at $\delta_l\approx$0.05 while $\delta_u$ was varied from -0.36 to -0.18. At the strongest negative $\delta_u$, no ELMs are observed in the $D_\alpha$ filterscope signal at the bottom left of Fig. 13. At a moderate $\delta_u$ of -0.3, characteristic dithering behavior is found, which is usually interpreted as proximity to the L-H transition. Conversely, at the weakest shaping ($\delta_u$= -0.18), distinct ELMs are observed. 



\subsection{\label{sec:advances}Recent Advances}

\begin{table*}[p] 
\caption{\label{tab:nt_tokamaks}Overview of international tokamaks exploring NT plasmas. Detailed are the historical start dates, shaping capabilities, auxiliary heating, plasma parameters, and plasma-facing components (PFCs) utilized in these NT studies.}
\begin{ruledtabular}
\begin{tabular}{lclccl}
\textbf{Tokamak} & \textbf{Start} & \textbf{Shaping} & \textbf{$P_{aux}$ [MW]} & \textbf{$I_{p}$ [MA], $V$ [m$^3$]} & \textbf{PFCs} \\
\colrule
TCV & 1993 & Strongest NT flexibility: $\delta \ge -0.6$ & $<3$ & $\le 0.5, \le 1$ & Graphite \\
DIII-D & 2016 & Flexible wall: $\delta \ge -0.6$ diverted (2023) & $\le 15$ & $\le 1.2, \le 14$ & Graphite \\
AUG & 2017 & Weak: $\delta \ge -0.2$ & $\le 13$ & $\le 0.8, \le 10$ & Tungsten \\
JET & 2023 & Weak: $\delta \ge -0.05$ & $\le 32$ & $\le 1.7, \le 75$ & Tungsten, Be \\
MAST-U & 2024 & Weak: $\delta \ge -0.15$ & $\le 3$ & $\le 0.6, \le 11$ & Graphite \\
\end{tabular}
\end{ruledtabular}
\end{table*}

NT studies have been accelerating across the globe, spearheaded by the five major facilities shown in Fig. 14 and detailed in Table 2. The early foundational work on TCV established the core physics of the NT shape. As highlighted in this table, despite its small plasma volume, TCV features a uniquely open vacuum vessel and a highly adaptable poloidal field coil system. This hardware combination continues to offer the strong NT shaping flexibility and allows for the most comparisons with similar PT shapes. 

Following the initial exploratory NT studies on \mbox{DIII-D}, new temporary tiles were installed on this device in 2023 to protect the outer wall and diagnostics. This critical upgrade enabled operations in a $\delta\approx-0.5$ diverted shape. \mbox{DIII-D} then conducted a month-long, goal-focused campaign to further explore NT and address reactor relevancy \cite{Thome_2024, Paz-Soldan_2024}; most of the \mbox{DIII-D} NT database is in this strongly NT diverted shape. 

Concurrently, AUG has continued its initial NT studies, pushing toward stronger shaping to avoid H-mode access \cite{Vanovac_2024,Vanovac_NF_submitted}, despite the geometric limitations of its vacuum vessel. Recently, \mbox{MAST-U} became the first spherical tokamak to operate in the NT shape \cite{Nelson_2024}. Finally, during JET's last campaign, the largest NT plasmas to date were produced, successfully remaining in L-mode despite up to 32 MW of injected heating power \cite{Labit_2024, dunne2026physics}. TCV, \mbox{DIII-D} and \mbox{MAST-U} all have carbon plasma facing components (PFCs), whereas AUG and JET both have metal PFCs. Additionally, initial studies to develop the shape are underway on WEST \cite{marinoni_dpp2025_west}, SMART, and KSTAR. It is worth noting that the majority of the tokamaks described here, with the exception of TCV and SMART, were natively designed for PT due to their poloidal coil placement, control systems, and vacuum vessel geometries, so developing high-power NT shapes is a significant engineering challenge.

The remainder of this subsection will evaluate recent NT experimental advances across the three pillars described in Sec. II: performance, exhaust, and robustness. The vast majority of NT data and publications originate from TCV and \mbox{DIII-D} since they have studied NT the longest and have achieved the strongest NT plasmas. This success is due to their vacuum vessel flexibility, the placement of poloidal coils on both the inboard and outboard side, and their robust control systems. Consequently, results from these two devices will feature prominently throughout the following discussion. 

\subsubsection{\label{sec:advances}Performance}

As introduced in Sec. IIA, evaluating global tokamak performance traditionally centers on two primary metrics: $\betan$ and $H_{98y,2}$. This section will begin by examining how NT plasmas perform against these benchmarks. Next, the discussion will focus on how current NT confinement data extrapolates to a reactor, followed by a brief description of fast-ion transport studies between NT and PT.

\paragraph{\label{para:Conf}\textbf{Confinement}}~\\
While weakly-shaped NT plasmas can access H-mode and exhibit ELMs, the focus of this paper is primarily on the reactor-relevant ELM-free NT scenario. The access conditions for this regime will be described in detail in Sec. III.B.2, and this scenario will generally be referred to as NT unless explicitly stated otherwise. NT is neither an L-mode nor an H-mode scenario; rather, it constitutes a distinct regime with characteristics of both L-mode and H-mode that will be further described in the remainder of this section. Consequently, directly comparing NT with a PT counterpart can be challenging, as the equivalent PT discharge typically is in the H-mode regime with the same applied power. Furthermore, much like how H-mode encompasses various scenarios, such as those with different ELM types and the advanced tokamak, the NT regime is not a monolith. While different types of NT regimes exist, they have not been fully explored due to the nascent state of NT studies.

Both \mbox{DIII-D} \cite{Austin, Marinoni2019, Marinoni_2021, Paz-Soldan_2024, Thome_2024} and TCV \cite{Coda_2022,Balestri_2025} robustly achieve $H_{98y,2}\gtrsim1$ in their NT plasmas, as shown in Fig. 15, across a wide range of edge safety factors ($q_{95}$). The \mbox{DIII-D} data [Fig. 15(b)] shown in this figure are exclusively from the 2023 campaign with the strong NT, but are consistent with previous results in both strong NT limited and weakly NT diverted shapes. AUG has also observed $H_{98y,2}\approx1$ in some plasmas without high amounts of fast ions\cite{Happel_2023}. Leveraging higher amounts of auxiliary heating, \mbox{DIII-D} routinely accesses $\betan>2.5$ and approaches an ideal limit of 3 \cite{Marinoni2019}, which is somewhat lower than the ideal limits typically observed in PT. In contrast, AUG \cite{Happel_2023, Vanovac_2024}, \mbox{MAST-U} \cite{Nelson_2024}, and JET \cite{Labit_2024} all predominantly find $H_{98y,2}<1$ at the weaker NT shapes dictated by their vacuum vessel limitations. However, it must be emphasized that there has been limited experimental runtime in all three tokamaks compared to \mbox{DIII-D} and TCV. It should also be noted that for spherical tokamaks, such as \mbox{MAST-U}, theoretical predictions indicate that the underlying transport mechanisms differ fundamentally from those at higher aspect ratios \cite{Marinoni_2024, Balestri_2024}. 
\begin{figure}
\includegraphics[width=1\linewidth, trim=3cm 0cm 3cm 0cm,
    clip] {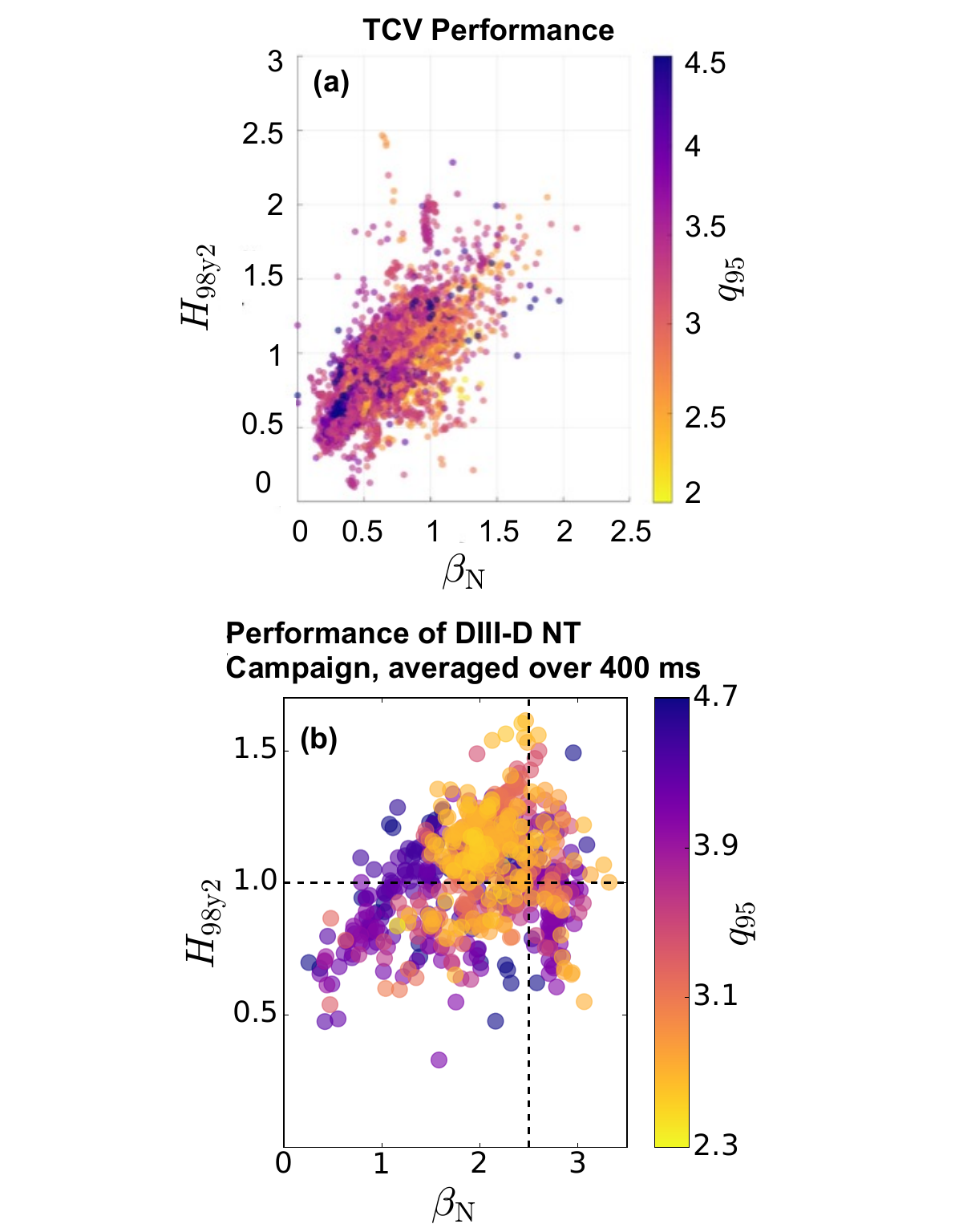}
\caption{\label{fig:figure13} Comparison of $\betan$ versus $H_{98y,2}$ for TCV (a), adapted from \cite{Balestri_2025}, and \mbox{DIII-D} (b) NT plasmas at various $q_{95}$, adapted from \cite{Thome_2024}. }

\end{figure}

Observations from \mbox{DIII-D} and TCV indicate that confinement improves with stronger NT, as illustrated by Fig. 16. The \mbox{DIII-D} data [Fig. 16(b)] are clustered at three NT values: $\delta\approx-0.5$ in the strong diverted NT shape, $\delta\approx-0.4$ in the limited NT shape, and $\delta\sim-0.1$~ in the weakly diverted NT shape. The maximum $H_{98y,2}$ factor increases as the magnitude of the NT increases. Due to limitations imposed by the poloidal coil system and the availability of wall armoring on \mbox{DIII-D}, it is not possible to dynamically sweep between the different shapes during a single experiment\cite{Thome_2024}. Consequently, \mbox{DIII-D} has limited ability to continuously scan the triangularity parameter space.  However, TCV is able to scan both the upper and lower triangularity independently in Ohmic, single-null diverted plasmas. These experiments indicate that TCV confinement is primarily driven by the triangularity of the non-X-point half of the plasma [Fig. 16(a)] \cite{coda_eps2023}. The confinement changes described here are likely tied to microinstability changes, as turbulence amplitude and type have also been found to change in NT plasmas as a function of $\delta$ \cite{Stewart_2025}. 

\begin{figure}
\includegraphics[width=1\linewidth, trim=0cm 0cm 3cm 0cm,
    clip] {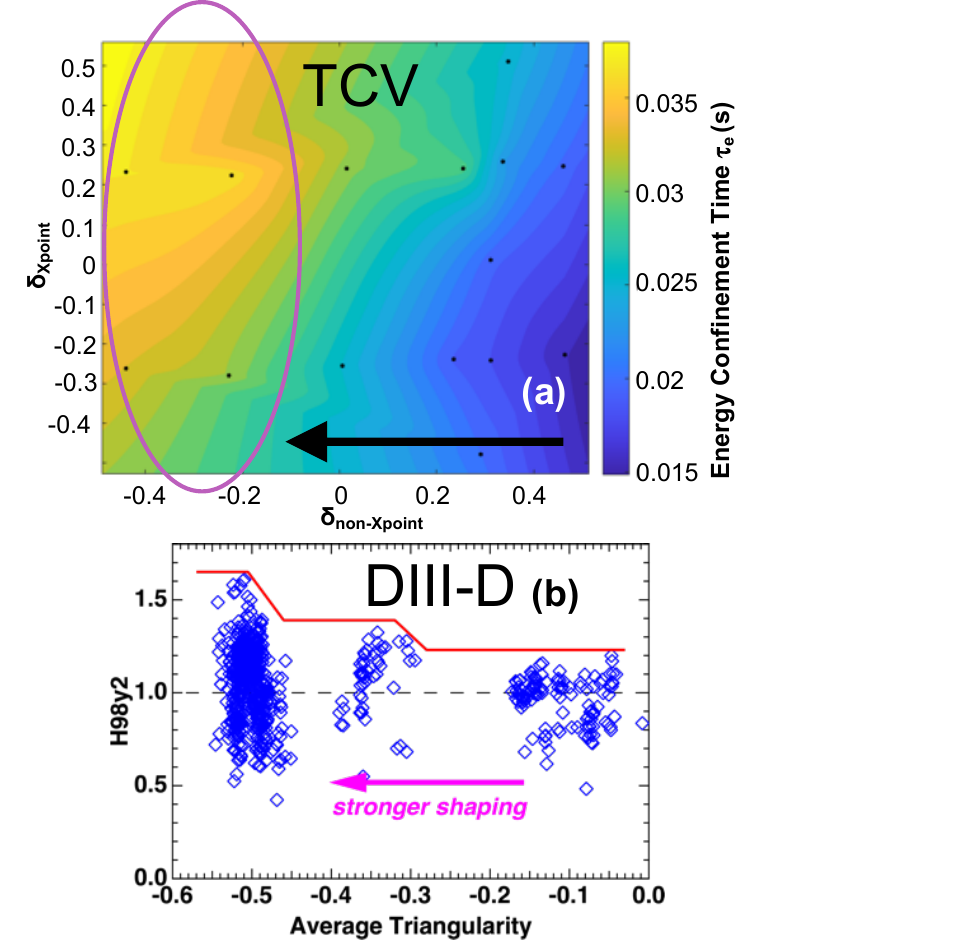}
\caption{\label{fig:figure13} Dependence of confinement on triangularity. (a): Comparison of the TCV energy confinement time $\tau_e$ on $\delta$ at in Ohmic plasmas varying the $\delta_l$ and $\delta_u$, courtesy of S. Coda, reproduced from \cite{coda_eps2023}. (b): $H_{98y,2}$ as a function of average triangularity in \mbox{DIII-D} NT plasmas with the maximum noted by the red line, adapted from \cite{Thome_2024}. }
\end{figure}

Future reactors will operate in parameter regimes that are fundamentally distinct from today's tokamaks: specifically featuring a lower normalized gyroradius, $\rho_*$, reduced rotation from lower externally injected torque, and different forms of heating and current drive. Extrapolating performance from our current machines to these devices is essential and relies on a rigorous synthesis of experimental data, theoretical frameworks, and advanced predictive modeling. To this end, higher-fidelity NT modeling is currently advancing for both the core and edge \cite{Marinoni_2024,Bielajew_2025,becoulet_submitted,ulbl_submitted, McClenaghan_2024, Hoffmann_2025, DiGiannatale_2024}. 

\begin{figure}
\includegraphics[width=1\linewidth, trim=0cm 0cm 0cm 0cm,
    clip] {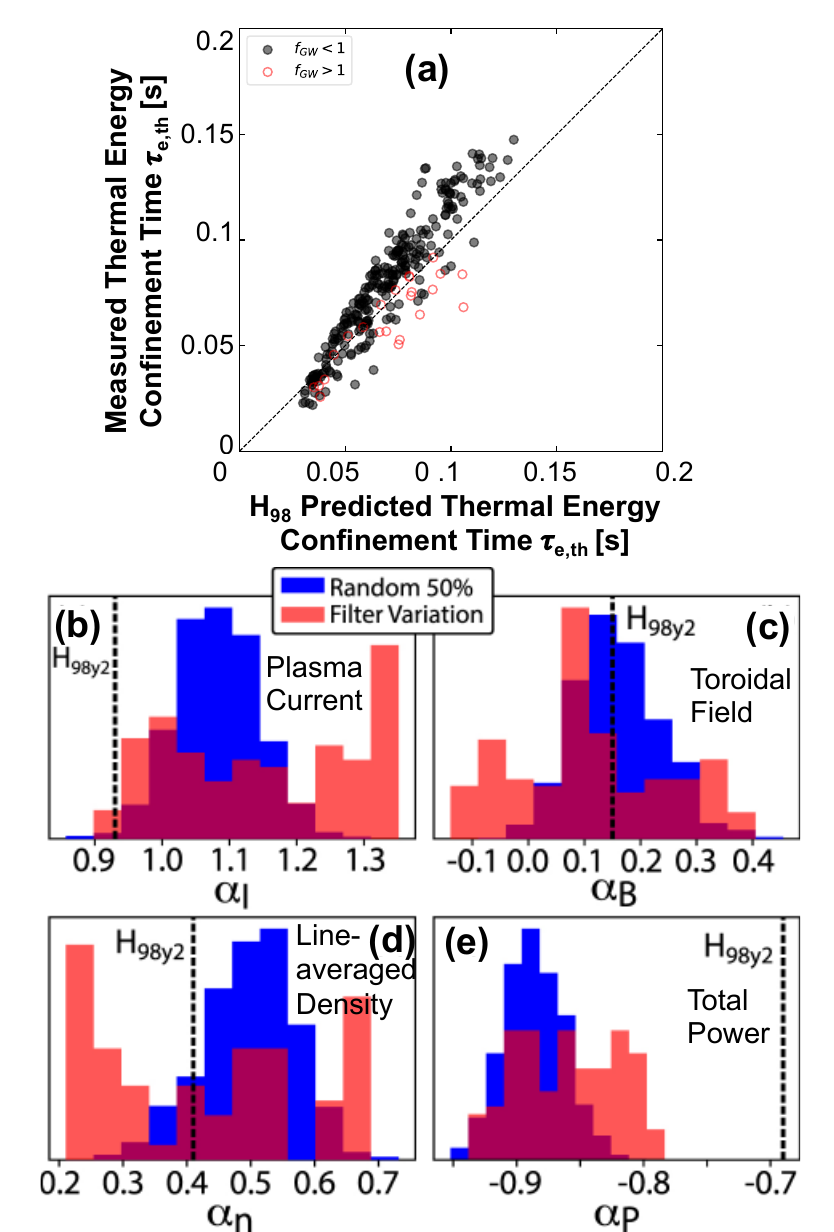}
\caption{\label{fig:figure18} Comparison of \mbox{DIII-D} confinement data with predictions from the $H_{98y,2}$ scaling. (a): Comparison of the measured thermal energy confinement time $\tau_{e_{th}}$ with that predicted by the $H_{98y,2}$ scaling at densities below (full black circles) and above (open red) the Greenwald fraction ($f_{GW}$). Comparison of the exponents of the engineering parameters plasma current (b), toroidal field (c), line-averaged density (d), and total power (e) in NT plasmas with the exponent of that parameter in $H_{98y,2}$ scaling indicated by the dashed line. Adapted from \cite{Paz-Soldan_2024}. }
\end{figure}

A common method for extrapolating performance from present-day devices to future reactors is to benchmark data against standard multi-machine scalings. Evaluating the \mbox{DIII-D} thermal energy confinement time against the standard PT H-mode $H_{98y,2}$ empirical scaling reveals a broad consistency,  as shown in Fig. 17(a). However, a dependence on the normalized plasma density, the Greenwald fraction ($f_{GW}$), is observed, with confinement degrading at higher $f_{GW}$. A comparison of the exponents of some of the engineering parameters in the $H_{98y,2}$ scaling is also provided [Fig. 17(b--d)]. Notably, the NT data exhibit a stronger dependence on plasma current [Fig. 17(b)] and a more significant power degradation [Fig. 17(e)] than in the PT scaling. Furthermore, \mbox{DIII-D} and TCV similarity experiments indicate that confinement falls between the Bohm and the more optimistic gyro-Bohm $\rho_*$ scalings \cite{marinoni_nf_submitted}. While this is inconsistent with the gyro-Bohm nature of the $H_{98y,2}$ scaling, it is important to note that these experiments spanned only a narrow range of $\rho_*$.

In modern tokamaks, the energy confinement time is typically measured in highly rotating plasmas driven by externally injected torque. While rotation is known to stabilize transport \cite{burrell1997}, fusion reactors are expected to operate with low external torque and, consequently, low rotation \cite{coliniter}. The dependence of the thermal energy confinement time ($\tau_{e_{th}}$) on torque was measured on \mbox{DIII-D} by varying the neutral beam power and torque alongside the ECRH power. In standard PT plasmas, confinement is typically found to degrade as rotation is decreased. Similar trends were observed in NT, particularly at low $q_{95}$ where $\tau_{e_{th}}$ decreases alongside torque. However, $\tau_{e_{th}}$ was found to not change at higher $q_{95}$ at lower injected torque \cite{Chrystal_2024}. Furthermore, nonlinear gyrokinetic simulations of \mbox{DIII-D} limited plasmas corroborate this dependence of confinement on rotation \cite{Marinoni_2024}. 

Ultimately, confidently extrapolating NT confinement to a future reactor requires expanding the empirical database across a broader range of engineering and physics parameters. While the foundational datasets from TCV and \mbox{DIII-D} are robust, they are inherently constrained by facility capabilities. TCV operates with relatively low auxiliary power, and the \mbox{DIII-D} NT 2023 campaign utilized mainly NBI-dominant heating since ECRH was marginally available at the time.  NBI introduces significant torque and core fueling—conditions that contrast sharply with the low-torque, RF- and alpha-dominated electron heating expected in a burning plasma. To fully validate predictive models for future power plants, the international community must acquire complementary NT confinement data across many tokamaks over a much larger $\rho_*$ space, operate with reactor-relevant metal walls, and utilize dominant RF heating on both existing and future devices. Advancing the empirical database in even one of these specific areas will serve as a critical step forward for reactor extrapolation.

\paragraph{\label{para:fast}\textbf{Fast-ion transport}}~\\
Fast-ion transport has not been studied in as much  detail as thermal transport, but so far indications are that NT fast-ion confinement is comparable or better than in PT. On \mbox{DIII-D}, a detailed fast-ion transport study was conducted in the limited shape with matched NT and PT plasmas exhibiting similar fast-ion transport levels \cite{VanZeeland_2019}. A similar comparison study on TCV in limited NT and PT shapes also determined that fast-ion confinement is agnostic to a sole change in $\delta$ \cite{Poley-Sanjun_2026}. Numerical simulations indicate that the NT shaping  may be a potential candidate to reduce fast ion transport due to unstable Alfv\'en eigenmodes in fusion reactors \cite{Ghai_2021}. Furthermore, recent numerical modeling of TCV plasmas reports a potential mitigation of Toroidal Alfv\'en Eigenmodes and a reduction in associated fast-ion losses at NT \cite{OyolaDominguez_Thesis2024}.

\subsubsection{\label{sec:advances}Exhaust}

Much like conventional PT tokamak plasmas, NT configurations must also successfully manage both steady-state and transient fluxes of both heat and particle exhaust. Addressing the transient fluxes first, the robust prevention of ELMs must be evaluated, as this eliminates one of the most severe sources of intermittent exhaust. Next, the discussion will pivot to steady-state exhaust challenges, specifically examining the power exhaust footprint through two primary metrics: the heat flux decay length ($\lambda_q$) and the accessibility of divertor detachment, including utilizing radiative mantle. Finally, an assessment of impurity confinement is presented. 

\paragraph{\label{para:ELMs}\textbf{ELMs}}~\\
Building upon the initial demonstration of NT plasmas achieving high performance and remaining ELM-free during the foundational NT experiments discussed in Section IIIA, the operational database has expanded significantly to include a wider array of both existing and new tokamaks, as summarized in Table 2.  Across these devices, plasmas at sufficiently strong NT inherently avoid transient exhaust from ELMs by not accessing H-mode. This robust, ELM-free behavior has been experimentally confirmed across a wide range of machines, including TCV \cite{Pochelon2012TCV, Coda_2022}, \mbox{DIII-D} \cite{oakprl}, AUG \cite{Vanovac_NF_submitted}, \mbox{MAST-U} \cite{Nelson_2024}, and JET \cite{Labit_2024, dunne2026physics}. The \mbox{DIII-D} NT operational space is illustrated in Fig. 18(a); at all $\delta<\delta\mathrm{_{crit}}$, no ELMs are observed, even with up to a factor of ten times more applied power than the predicted L-H power threshold. While all diverted \mbox{DIII-D} plasmas in this database were LSN with the X-point at the bottom, recent findings indicate that the upper triangularity, the non X-point triangularity in this case, may contribute more significantly to the avoidance of ELMs.
\begin{figure}
\includegraphics[width=1\linewidth, trim=0cm 0cm 0cm 0cm,
    clip] {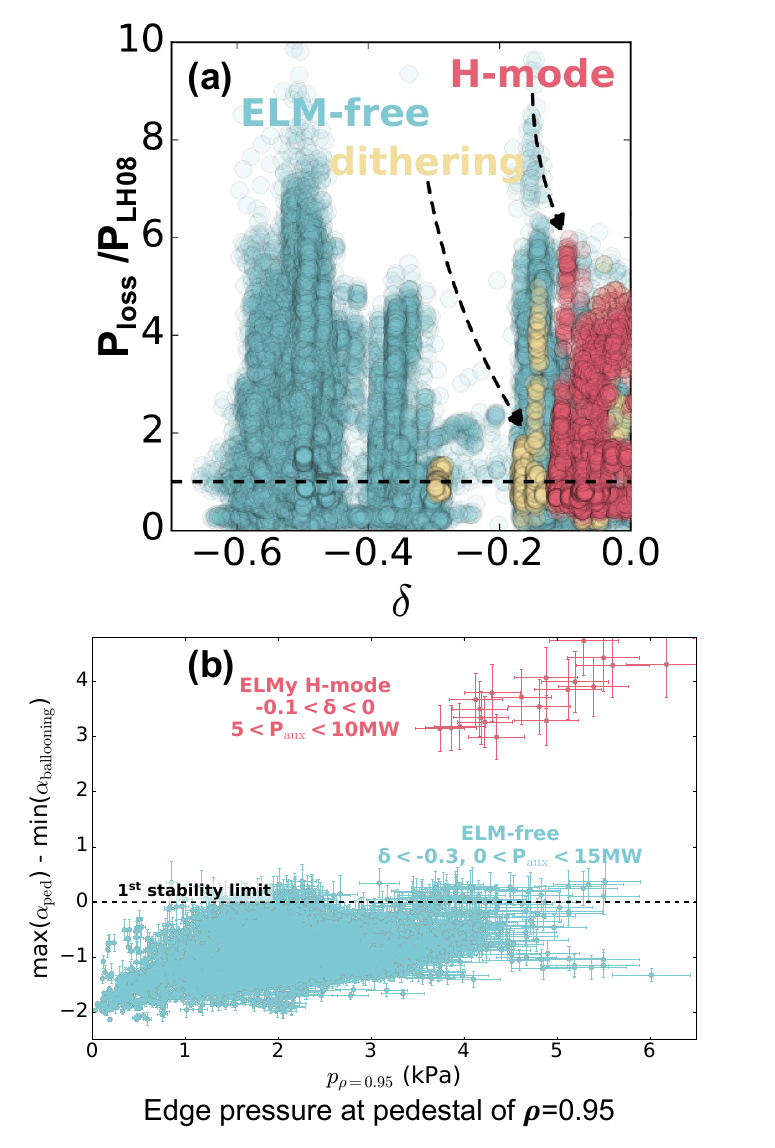}
\caption{\label{fig:figure18} \mbox{DIII-D} NT ELM operating space through 2023. (a): The H-mode power threshold $P_{loss}$/$P_{LH08}$ as a function of $\delta$. ELMy H-modes are colored in red, dithering periods in yellow, and ELM-free periods in blue. Adapted from \cite{oakprl}. (b): Edge pressure at the pedestal compared to the ballooning stability limit.  Some select ELMy NT plasmas are also shown. Adapted from \cite{oakprl}. }

\end{figure}

Importantly, this ELM avoidance is not a transient phase or a highly sensitive tuned condition, as is often the case with PT ELM-suppression techniques. During the dedicated \mbox{DIII-D} NT campaign, ELM-free high-performance NT plasmas were maintained reliably discharge after discharge, day after day, throughout the entirety of the campaign, despite numerous aggressive and deliberate efforts to trigger H-mode. However, at $\delta>\delta\mathrm{_{crit}}$ on \mbox{DIII-D}, most plasmas transition into H-mode and have ELMs, sometimes even with applied power below that predicted by the L-H threshold scaling. At $\delta$ around $\delta\mathrm{_{crit}}$ a dithering behavior between L-mode and H-mode is observed. The remarkable resilience of this regime was perhaps most starkly demonstrated during recent JET campaigns, where the plasma did not access H-mode despite the application of 32 MW of injected auxiliary power \cite{Labit_2024, dunne2026physics}.

In PT plasmas, H-mode is usually accessed in the second stability ballooning regime but can on rare occasions be entered in the first stability ballooning regime. Access to the ballooning second stability regime is blocked when $\delta<\delta\mathrm{_{crit}}$ on TCV \cite{Pochelon2012TCV, Medvedev_2015}, \mbox{DIII-D} \cite{oakprl}, AUG\cite{Vanovac_NF_submitted} and JET \cite{Labit_2024, dunne2026physics}. This is illustrated in Fig. 18(b), where the edge pressure in the pedestal is plotted for all of the \mbox{DIII-D} NT 2023 campaign discharges and is shown to be constrained by the first stability limit within experimental error bars. In contrast, discharges with weaker NT retain access to the second stability regime, H-mode, and ELMs. As can be seen from this figure,  the ideal ballooning (infinite-n) boundary clamps the pedestal height below the ELM limit. In NT plasmas, there is only a small edge pedestal and, as can be seen from this figure, there is some variation in the pedestal height, which is an active area of investigation \cite{Nelson_ppcf}. Spherical tokamaks such as \mbox{MAST-U}, can more readily access H-mode in the first ballooning regime. Thus, when the plasma has $\delta<\delta\mathrm{_{crit}}$, access to second stability in MAST-U is blocked but small ELMs were still observed. Eventually an ELM-free regime on MAST-U is observed as the plasma becomes more strongly NT. Therefore, unlike at higher aspect ratio, where the closure of second stability alone determines access to ELM-free NT plasmas, the ELM operating space in spherical tokamak NT plasmas is influenced by additional physics mechanisms that are not yet fully captured by standard high-aspect-ratio models \cite{Nelson_2024}.

\begin{figure}
\includegraphics[width=1.\linewidth, trim=0cm 0cm 0cm 0cm,
    clip] {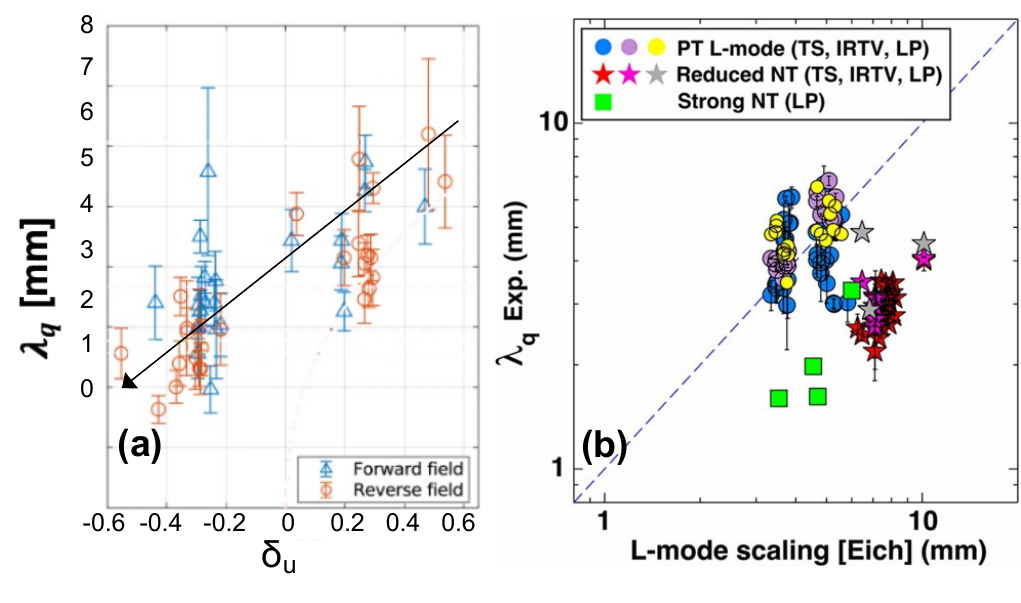}
\caption{\label{fig:figure18} $\lambda_q$ in NT plasmas on TCV as a function of $\delta$ (a), adapted from \cite{Morgan_2025}, and \mbox{DIII-D} (b) compared to the standard L-mode $\lambda_q$ Eich scaling, reproduced from \cite{Scotti_2025}. }

\end{figure}

\paragraph{\label{para:hf}\textbf{Steady-state exhaust}}~\\
With ELMs robustly prevented, the focus shifts to steady-state exhaust. In tokamaks, the narrow scrape-off layer sets the baseline steady-state exhaust challenge, characterized by the heat flux decay length ($\lambda_q$) in tokamaks. Excess particle and heat fluxes over a small wetted area can lead to component damage and impurity influxes, as described in Section IIB. Typically, PT H-mode has a narrower $\lambda_q$ than PT L-mode due to reduced turbulence levels. 

The dependence of $\lambda_q$ on $\delta$ has been studied on both TCV and \mbox{DIII-D}, as shown in Fig. 19. On TCV, it was studied in Ohmic plasmas with the magnetic field in both directions and found to decrease with NT [Fig. 19(a)]\cite{Morgan_2025}.  On \mbox{DIII-D}, $\lambda_q$ was studied in both the strong NT and weakly NT diverted shapes and compared with the standard L-mode $\lambda_q$ Eich scaling [Fig. 19(b)]\cite{Scotti_2025}. In both DIII-D shapes, $\lambda_q$ is narrower than that predicted by the L-mode scaling and also that measured in \mbox{DIII-D} PT L-mode plasmas as it is closer to that predicted by the standard H-mode $\lambda_q$ Eich scaling. Consistent with TCV results, $\lambda_q$ is shorter in the stronger NT shape than the weaker diverted shape. The narrower $\lambda_q$ is consistent with the reduced turbulence levels seen at strong NT shaping. 

To prevent damage and impurity influxes, the exhaust plasma must be cooled before it reaches the reactor’s divertor target. Power can be radiated isotropically to spread the exhaust energy over a much larger area than conducting it all to the strike points, which form a narrow ring. This radiation can be localized in the divertor region (radiative divertor or detachment), the confined plasma (radiative mantle), or both.  Both techniques are commonly used in PT plasmas, are considered essential for future reactors, and have been studied in NT plasmas, as will be discussed next.

\begin{figure}
\includegraphics[width=1.\linewidth, trim=0cm 0cm 0cm 0cm,
    clip] {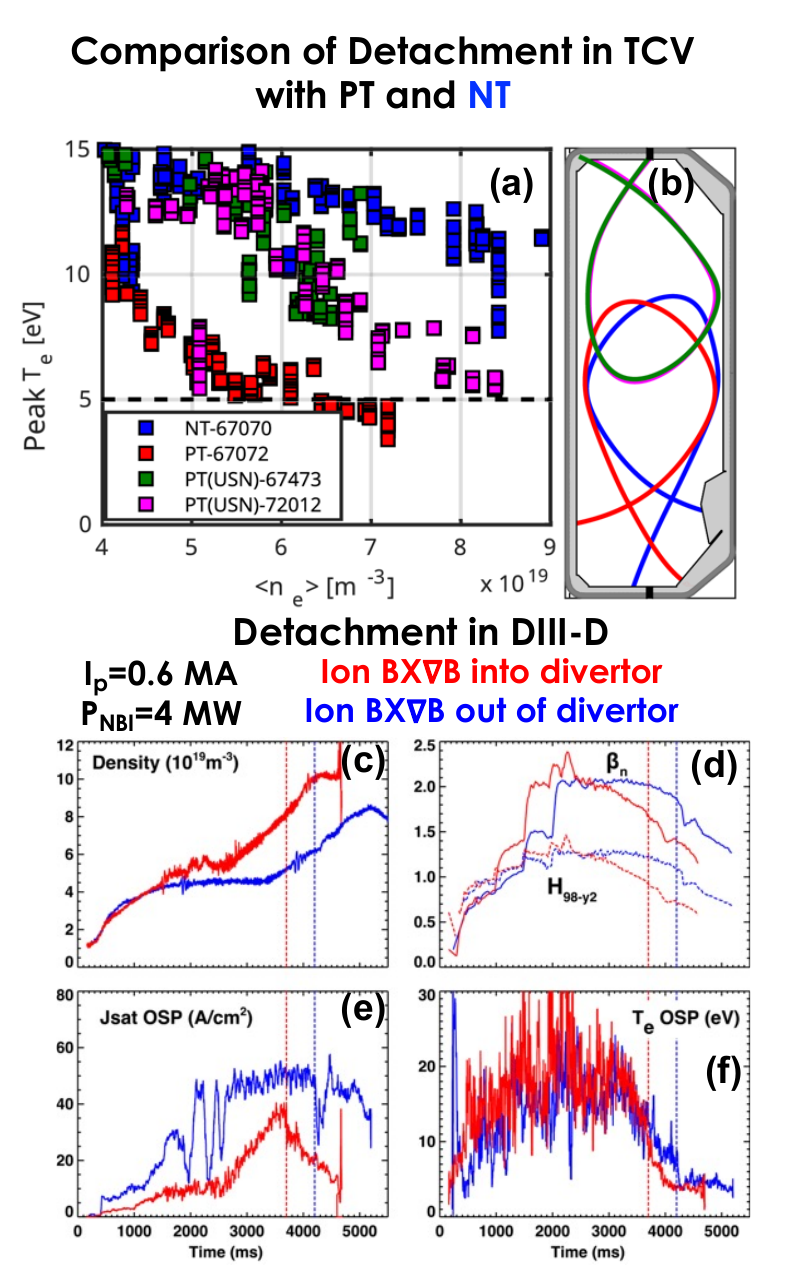}
\caption{\label{fig:figure18} Detachment in NT plasmas. Comparison of detachment in PT and NT on TCV with peak electron temperature $T_e$ vs density (a) and (b) respective shapes, reproduced from \cite{Fevrier_2024}. Achievement of detachment on \mbox{DIII-D} in both favorable ion gradB (red) and unfavorable (blue) drift directions with density (c), confinement parameters $\beta_N$ and $H_{98y,2}$ (d), saturation current on the Langmuir probes (e) and $T_e$ at the outer strike point (OSP) (f), reproduced from \cite{Scotti_2025}. }

\end{figure}
Detachment occurs when impurity radiation and recombination processes reduce heat and particle fluxes to the material surface; it is achieved by increasing radiative exhaust via impurity seeding or upstream density increase, commonly exploiting intrinsic carbon radiation on DIII-D or TCV. Thus, the detachment density is usually an important figure of merit, defined when the plasma cools to a certain electron temperature, usually 2 eV. Detachment has been achieved in NT plasmas on both \mbox{DIII-D}\cite{Scotti_2024,Scotti_2025} and TCV\cite{Fevrier_2024}, but it requires higher densities than in PT plasmas, as can be seen in Fig. 20 (a) on TCV with the comparison of detachment in PT and NT shapes [Fig. 20(b)]. TCV requires the injection of extrinsic impurities (N) to detach. While \mbox{DIII-D} is able to detach with intrinsic carbon radiation, extrinsic impurities lower the detachment density. Detachment densities on \mbox{DIII-D} were typically higher than 90\% of the Greenwald fraction. However, as can be seen on \mbox{DIII-D}, as the density [Fig. 20(c)] increases, the confinement starts to drop, as indicated by decreases in $\betan$ and $H_{98y,2}$ [Fig. 20(d)], in agreement with Fig. 17(a). This detachment difficulty is due to the NT geometry (shorter poloidal and parallel divertor lengths) and reduced radial transport ($\lambda_q$) when compared to PT L-mode. Detachment is  easier at reduced NT shapes \cite{Scotti_2025}. The ability to reproduce detachment behavior with fluid codes provides confidence in extrapolating dissipative divertor solutions \cite{Zhao_2025, Lore_SOLPS, Tonello_2024}.

Increasing the radiation in the confined plasma and creating a radiative mantle was performed on \mbox{DIII-D} in the 2023 campaign shape by injecting neon, argon, and krypton gases for core seeding, either individually or in combination with nitrogen for edge seeding. A comparison of the achieved $\beta_N$ on \mbox{DIII-D} with these various gases and the radiative core fraction $f_{rad_{core}}$ is shown in Fig. 21.  $f_{rad_{core}}$ was increased from 0.2 in the unseeded reference case to 0.45--0.55, which is comparable to the best achieved in PT scenarios. At the same time, the increased total radiated power fraction went from 0.5 to 0.8. A key advantage of this scenario in NT plasmas is that it is not limited by the need to stay above the L-H power threshold \cite{Thome_2024, casali}. This experiment tested controllability and impurity accumulation limits and was limited by the lack of active pumping in this shape \cite{eldon}. 

\begin{figure}
\includegraphics[width=1.\linewidth, trim=0cm 0cm 0cm 0cm,
    clip] {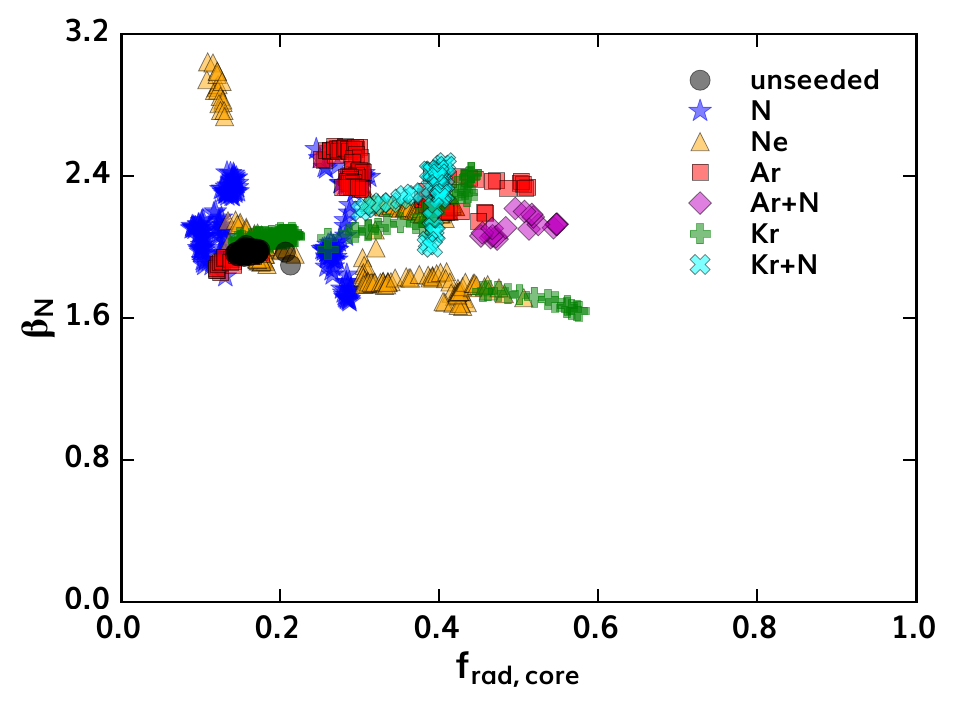}
\caption{\label{fig:figure21} \mbox{DIII-D} radiative mantle operating space with $\beta_N$ versus fraction of core radiated power, reproduced from \cite{Thome_2024}. }
\end{figure}

\paragraph{\label{para:hf}\textbf{Impurity Confinement}}~\\
Impurity transport has been studied on both \mbox{DIII-D} and TCV. Generally, \mbox{DIII-D} NT plasmas have low $Z_{\mathrm{eff}}\sim2$ with graphite plasma components in the various NT shapes \cite{Marinoni_2021, Sciortino_2022, Thome_2024}, which is lower than that seen in other high-performance ELM-free scenarios \cite{Paz-Soldan_2021}. Core impurity peaking is observed for high-Z impurities, similar to PT plasmas. TCV sees higher carbon density in limited NT plasmas than comparable PT plasmas \cite{Bagnato_2024}. The impurity confinement time $\tau\mathrm{_{imp}} $ was measured on \mbox{DIII-D} using laser blow-off injection. Impurity transport is often evaluated by comparing the impurity confinement time with the energy confinement time, as shown in Fig. 22 for both NT plasmas and PT H-mode plasmas. As can be seen in this figure, both $\tau\mathrm{_{imp}}/\tau_e$ and $Z_{\mathrm{eff}}$ are typically lower in NT plasmas than in PT plasmas, leading to reduced impurity accumulation.

\begin{figure}
\includegraphics[width=1.\linewidth, trim=0cm 0cm 0cm 0cm,
    clip] {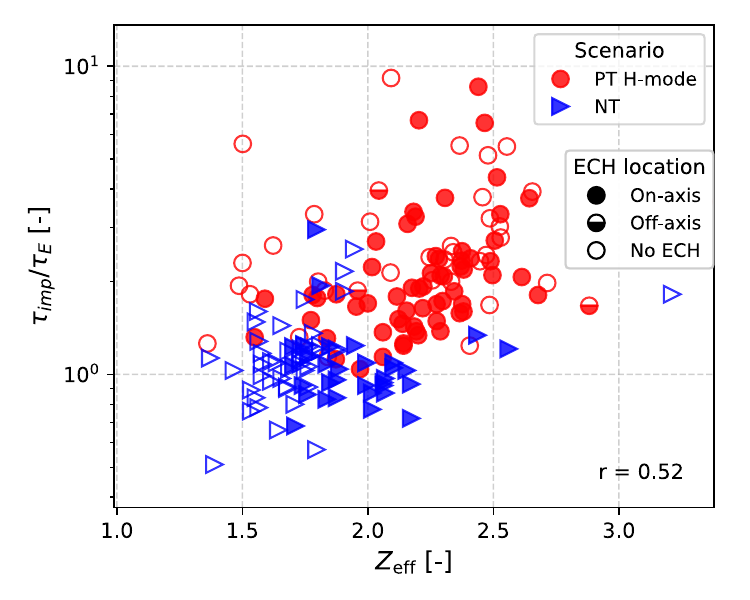}
\caption{\label{fig:figure21} \mbox{DIII-D} studies of the impurity confinement time using medium and high-Z impurities. Comparison of impurity confinement time to energy confinement time $\tau\mathrm{_{imp}}/\tau_e$ as a function of $Z_{\mathrm{eff}}$ for NT (blue triangles) and PT H-mode (red circle) plasmas with and without ECH injection. }
\end{figure}

\subsubsection{\label{sec:advances}Robustness}
\begin{figure}
\includegraphics[width=1.\linewidth, trim=0cm 0cm 0cm 0cm,
    clip] {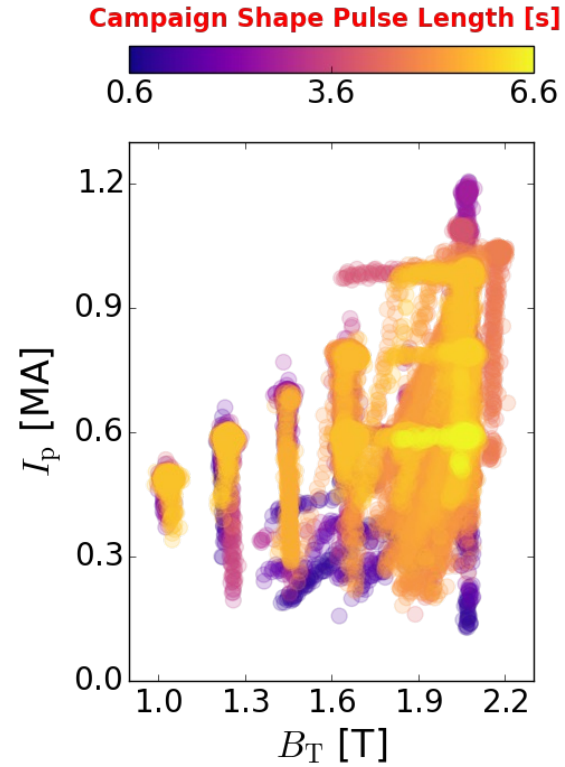}
\caption{\label{fig:figure21} \mbox{DIII-D} NT campaign operational space with $I\mathrm{_p}$ versus $B\mathrm{_T}$ with pulse length noted, reproduced from \cite{Thome_2024}.}
\end{figure}

NT plasmas have demonstrated excellent reproducibility, operating at high performance over a wide range of shapes and parameters on both TCV\cite{Coda_2022} and \mbox{DIII-D}\cite{Thome_2024,Paz-Soldan_2024}. Crucially, accessing this robust operating space does not require highly specialized wall conditioning, precise heating trajectories, carefully tuned plasma profiles, or complex tuning of 3D magnetic fields used for ELM suppression. Bypassing these specialized and fragile recipes, which are typically required for both accessing and maintaining proposed reactor-relevant PT H-mode scenarios, offers significant operational simplicity and reliability for future fusion reactors. Instead, the only limit to remaining ELM-free in NT is to stay below a machine's critical triangularity. Since this is purely a geometric requirement and a general understanding of the physics dependencies is known \cite{Nelson_2022}, the machine triangularity can be straightforwardly incorporated into the initial design phase of a reactor, eliminating the need for complex operational workarounds. What follows is a description of the wide operational space and discharge reproducibility in NT. 

\paragraph{\label{para:hf}\textbf{Wide Operational Space}}~\\
An example of the wide range of shapes achieved in NT plasmas for the major NT experiments worldwide is shown in Fig. 14. In this figure, TCV, which has the highest shape flexibility, has demonstrated LSN, upper-single-null and double-null diverted plasmas with a wide array of different upper and lower triangularities that are all robustly MHD-stable in Ohmic plasmas \cite{Coda_2022}. Similarly, DIII-D has three main NT configurations: an inner-wall-limited shape with $\delta=$-0.4\cite{Austin}, a weaker diverted shape with $\delta=-0.2$\cite{Marinoni_2021}, and the LSN diverted campaign shape with $\delta\approx-0.5$\cite{Thome_2024}. Developing and controlling these high-power NT shapes on DIII-D took significant experimental and control simulation time to realize on a machine optimized for PT. 
\begin{figure}
\includegraphics[width=1.\linewidth, trim=0cm 0cm 0cm 0cm,
    clip] {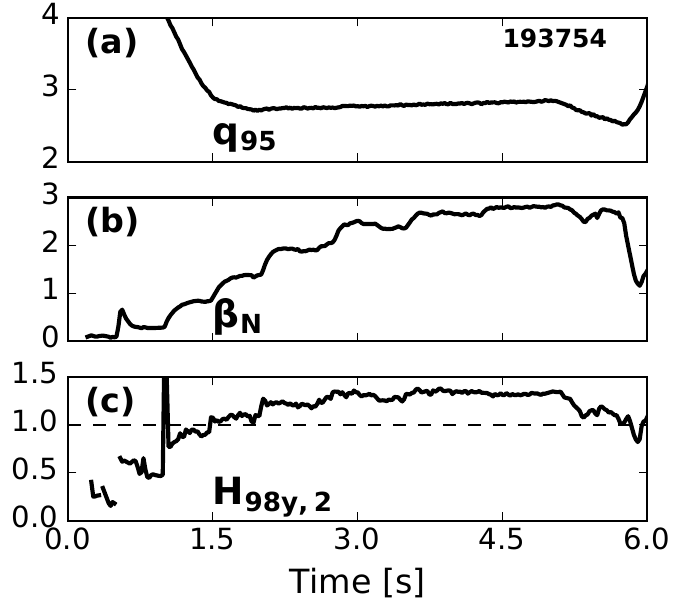}
\caption{\label{fig:figure21}  Low $q\mathrm{_{95}}$ high-performance plasma on \mbox{DIII-D} with $q\mathrm{_{95}}$ (a), $\beta_N$ (b), and $H_{98y,2}$ (c).}
\end{figure}

Utilizing the diverted NT campaign shape, a wide range of parameter space was accessed on DIII-D as shown in Fig. 23, with $I_p = 0.5\text{--}1.2$ MA and $|B_T| = 1.0\text{--}2.2$ T. These configurations were sustained over multi-second pulse lengths extending up to $\tau_{\mathrm{pulse}} \sim 6.5$~s across a wide range of safety factors ($q_{95} = 2.4\text{--}7$) and auxiliary heating powers up to $P_{\mathrm{tot}} \sim 15$~MW \cite{Thome_2024}. 

The NT configuration offers greater operational flexibility than conventional PT scenarios, particularly in parameter spaces critical for reactor performance. Both TCV\cite{Coda_2022} and \mbox{DIII-D}\cite{Thome_2024,Paz-Soldan_2024} are able to operate robustly at a low edge safety factor $q\mathrm{_{95}}<3$, a regime that typically triggers severe MHD instabilities and disruptions in PT H-mode plasmas. An example of accessing low $q\mathrm{_{95}}$ and maintaining high performance on \mbox{DIII-D} is shown in Fig. 24. 

While the relative disruption probability between PT and NT at low $q_{95}$ remains an open question, operating here opens up a possible new optimization space for an FPP by safely enabling higher plasma current at a given magnetic field. In particular, the inherent absence of ELMs removes a primary disruption trigger from this regime. To date, no major differences in disruptivity have been observed in NT compared to PT. Nevertheless, characterizing disruption behavior and mitigation requirements remains an essential area of ongoing experimental study for NT plasmas.

Furthermore, NT plasmas exhibit a remarkably high density limit, exceeding the Greenwald density limit using only edge gas puffing. Recent studies on \mbox{DIII-D} have demonstrated sustained, non-disruptive operation at Greenwald fractions up to $f\mathrm{_{GW}}\approx2$ \cite{Sauter_2025,Hong_2026} while the edge density remains generally below the Greenwald density, a regime typically inaccessible in the majority of PT H-mode scenarios. This capability is particularly interesting for reactor designs, as it allows for the high densities required for divertor detachment while simultaneously maximizing core fusion power density. 

\paragraph{\label{para:hf}\textbf{Reproducibility}}~\\
PT H-mode plasmas often exhibit marginal stability against MHD or edge instabilities, making very high-performance "trophy" discharges difficult to reproduce.  In plasmas with sufficiently strong NT, the elimination of ELMs improves the plasma stability and reproducibility. While these NT plasmas still have MHD instabilities\cite{Boyes_2023,Boyes_2024}
and lower overall predicted stability limits, they  have also demonstrated excellent discharge-to-discharge consistency at moderate normalized beta ($\beta_N\approx 2-3$) on \mbox{DIII-D}. 
\begin{figure}[t]
\includegraphics[width=0.9\linewidth, trim=0cm 0cm 0cm 0cm,
    clip] {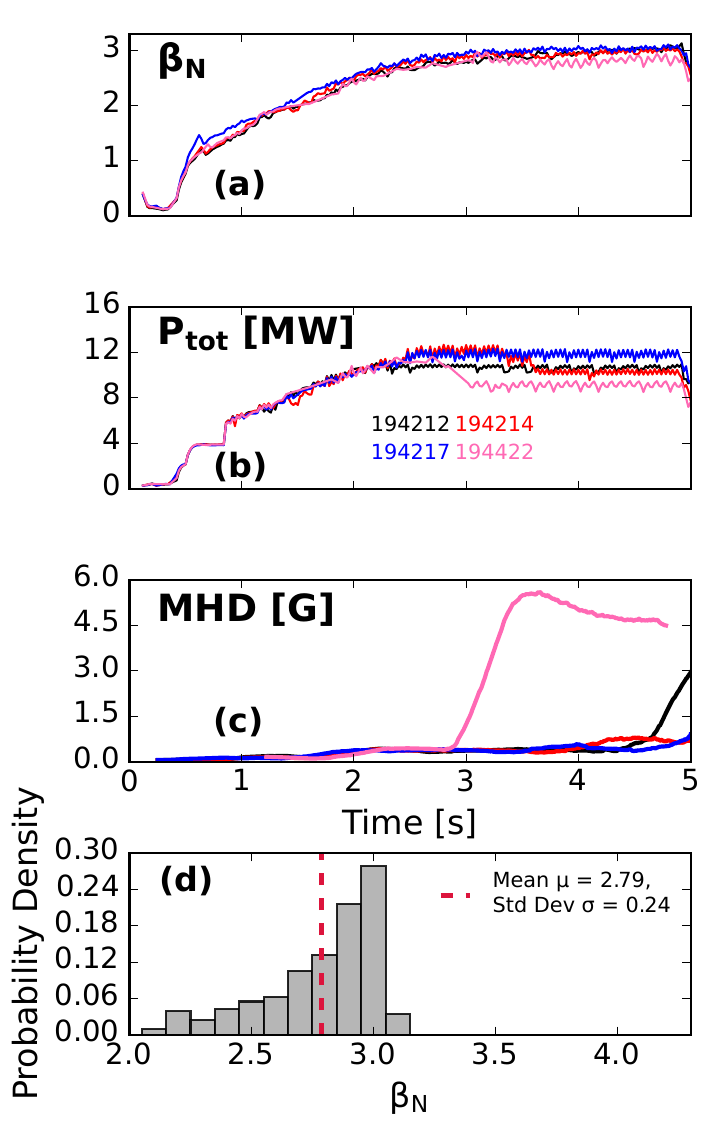}
\caption{\label{fig:figure7} Reproducibility of a moderate-beta NT \mbox{DIII-D} discharge compared to other discharges during that half run day with equivalent applied power: (a) $\betan$, (b) total auxiliary applied power, (c) $n=1$ MHD fluctuation amplitude and (d) the PDF of the operating distribution from these plasmas, where $P_\mathrm{tot}\geq$9 MW.   }
\end{figure}

A similar analysis as that performed for the very high performance PT H-mode half-day DIII-D experiment shown in Fig. 7 was conducted for a NT half-day experiment during the 2023 \mbox{DIII-D}  campaign that focused on creating stable moderate $\betan$ plasmas, as shown in Fig. 25. Repeat NT discharges on \mbox{DIII-D} are reproducible despite small changes in heating power [Fig. 25(b)] or MHD activity [Fig. 25(c)], and they notably do not require the precise tuning typically needed for high-performance PT scenarios.

To compare the PT and NT half-day experimental sessions, discharges with insufficient heating (less than 12 MW for PT and 9 MW for NT) or active day-of development tuning were filtered out of the comparative datasets, leaving 10 relevant discharges for the PT experiment and 4 for the NT experiment. Although the PT session benefited from a high-performance reference discharge to start from, the lower shot count for the NT case directly reflects that it lacked an established reference discharge, requiring extensive day-of development to establish its heating trajectories, alongside less coincidental hardware availability during that session. Looking at the performance, the PT half-run day achieved a higher mean $\beta_N$ of 3.16 compared to the NT mean $\beta_N$ of 2.79, but the PT experiment had a much higher standard deviation of 0.54 compared to the NT experiment of 0.24. This difference is clearly reflected in the structure of their probability density functions; the PT distribution is remarkably broad and flat due to the large shot-to-shot scatter [Fig. 7(d)], whereas the NT distribution forms a highly localized peak concentrated toward its upper operational limit [Fig. 25(d)].  This tight clustering demonstrates that NT plasmas populate a wide, reliable operating window at moderate $\beta_N$ that could serve as a  prerequisite for achieving high capacity factors necessary for an economically viable fusion reactor.

\section{\label{sec:futurepath}Future Work and Path to Pilot Plant}

As demonstrated by the experimental results in the previous section, NT is a promising route to sustainable tokamak reactors. This potential is reflected in the worldwide acceleration of NT studies in the past few years, which have made significant progress against the three pillars described in Sec. II:
\begin{itemize}
    \item \textbf{Performance}:  High confinement ($H_{98y,2}\gtrsim1$) has been achieved in strong NT on both \mbox{DIII-D} and TCV, with \mbox{DIII-D} accessing $\beta_N>$2.5. However, further data is required to understand reactor confinement extrapolation, particularly regarding RF-dominant heating and operation with metal walls. 
    \item \textbf{Exhaust}: ELM-free operation is accessible at sufficiently strong NT, with heat flux widths ($\lambda_q$) close to H-mode values. While  detachment is achievable, it occurs at higher densities than PT plasmas, which is explained by fluid codes. Additionally, NT plasmas have demonstrated that strong radiation in the confined plasmas is  compatible with this regime, and separately that there is low observed core impurity retention. 
    \item \textbf{Robustness}: A major advantage of this  regime is the large operating space of reproducible plasmas. This consistency allows for reliable access to both very high Greenwald fractions and low $q_{95}$ without the instabilities typical of high-performance PT discharges. 
\end{itemize}
Sec. IVA outlines the remaining physics questions that need to be resolved before transitioning tokamak reactors fully to the NT path, along with the planned future work on both existing and proposed machines to address them. Sect. IVB discusses the unique advantages of an NT-based pilot plant, including current design studies and the growing interest from the commercial fusion sector.   
\subsection{\label{sec:future}Future Work}

There are three outstanding questions from the worldwide NT work that motivate further exploration of NT as a reactor scenario: 
\begin{enumerate}
  \item \textbf{How does confinement scale to reactors?} As previously discussed in Sec. III B1, the missing pieces that need to be addressed for this question are the dependencies on power, plasma rotation, heating methods, and wall material. \mbox{DIII-D} confinement results indicate a strong degradation with power and a slight dependence on rotation. While multi-machine studies have been performed between \mbox{DIII-D} and TCV, further studies are needed to determine the $\rho_*$ scaling to extrapolate to reactors. Note, $\rho_*$ is predominantly a core-focused parameter and edge conditions are also important. Additionally, the majority of studies to date have been performed with dominant NBI heating on \mbox{DIII-D} with significant rotation, or plasmas with relatively low auxiliary power on TCV. Since reactors will be dominantly RF heated without external torque injection, studies in these regimes are essential. Finally, both \mbox{DIII-D} and TCV have carbon walls and there have been limited studies on machines with metal walls, which are the planned PFCs for a reactor. 
  \item \textbf{What is the optimal $\delta$ for the core and edge?} Confinement has been demonstrated to increase on both \mbox{DIII-D} and TCV with stronger NT. However, observations on both devices found that $\lambda_q$ becomes narrower at stronger NT. Identifying the optimal shaping to balance core performance with manageable divertor heat loads is therefore a critical priority. 
  \item \textbf{Can the core-edge integration be improved with a NT optimized closed pumped divertor? } So far, studies on both TCV and \mbox{DIII-D} have been performed in plasmas with an open unpumped divertor. Both devices have found that a very high density is required to detach, higher than PT, and \mbox{DIII-D} results show that confinement degrades at these high densities. Future work must determine if a closed divertor can facilitate detachment at lower densities and whether private flux region pumping can improve particle control, all while maintaining high core confinement.  
\end{enumerate}

Future NT upgrades and dedicated runtime are being planned across the existing NT tokamaks worldwide. \mbox{MAST-U}, WEST\cite{marinoni_dpp2025_west}, KSTAR, and SMART\cite{Cruz-Zabala_2024} will continue to develop their scenarios and extend them to stronger shaping, higher plasma current, and/or the increased auxiliary heating and current drive. AUG will continue to explore its recent no-ELM scenario. TCV will have more heating and current drive and has a highly flexible wall so the baffling can easily be changed. On \mbox{DIII-D}, a proposed upgrade featuring a closed pumped divertor, shown in Fig. 26, has been designed to enable scenarios at high confinement with a dissipative divertor to directly address these outstanding questions. 

\begin{figure}
\includegraphics[width=0.7\linewidth, trim=0cm 0cm 0cm 0cm,
    clip] {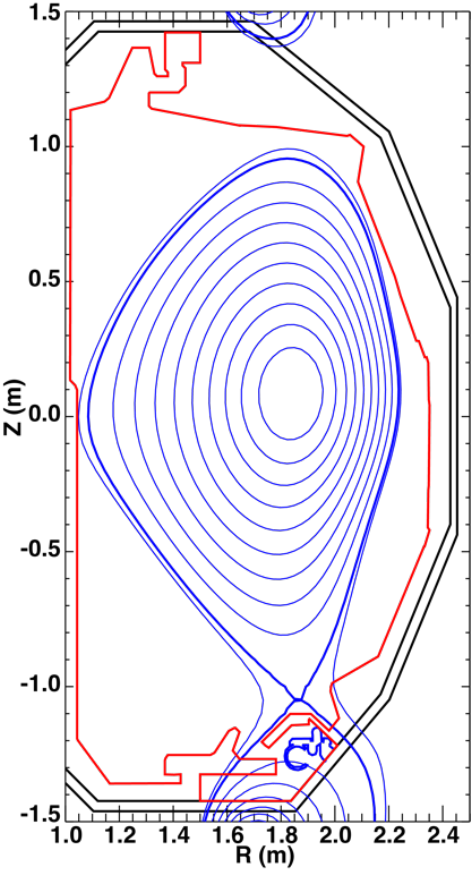}
\caption{\label{fig:figure7} Proposed \mbox{DIII-D} NT upgrade with a closed pumped divertor. }
\end{figure}

\subsection{\label{sec:path}Path to Pilot Plant}

There are several physics and engineering advantages of NT in tokamak FPP design that could directly lead to economic benefits:

\begin{itemize}
  \item \textbf{Simplified reactor core:} High performance with passive ELM stability removes dangerous transient exhaust, eliminating the need for ELM control. This obviates the requirement for complex and costly transient exhaust control systems, such as 3D magnetic coils or pellet injection, with no reduction in confinement. There is also effective impurity exhaust and a wide operational space of reproducible plasmas. 
   \item \textbf{No L-H threshold:} The absence of an L-H threshold allows for a very high core radiation fraction without the risk of dropping out of H-mode, potentially easing heat load requirements on the first wall. 
    \item \textbf{Exhaust geometry:} Placing the X-point on the outboard side provides a large exhaust area ($2\pi R_{div}$) and a more accessible divertor region, as illustrated by the comparison between an NT and PT plasma in Fig. 27. 
\end{itemize}

\begin{figure}
\includegraphics[width=1\linewidth, trim=0cm 0cm 0cm 0cm,
    clip] {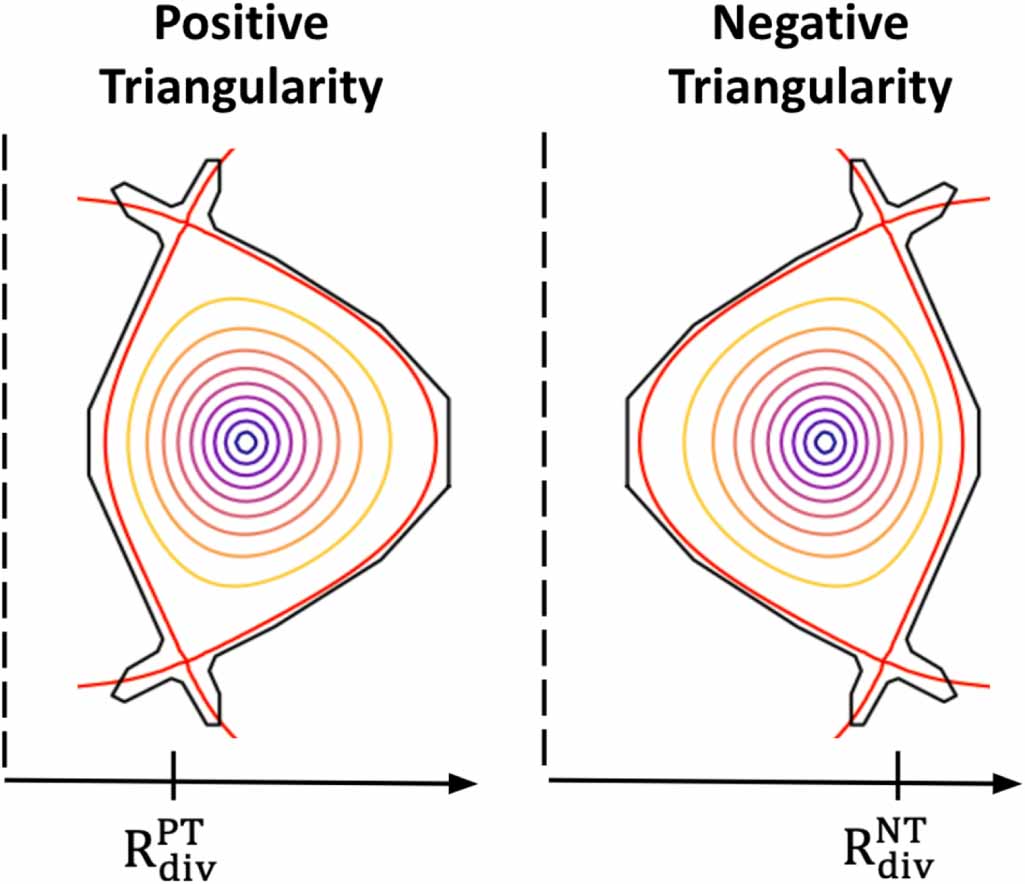}
\caption{\label{fig:figure7} Comparison of PT (left) and NT (right) plasmas, illustrating difference in divertor location, reproduced from \cite{Miller_2024}.}
\end{figure}
\begin{figure*}
\includegraphics[width=0.9\linewidth, trim=0cm 0cm 0cm 0cm,
    clip] {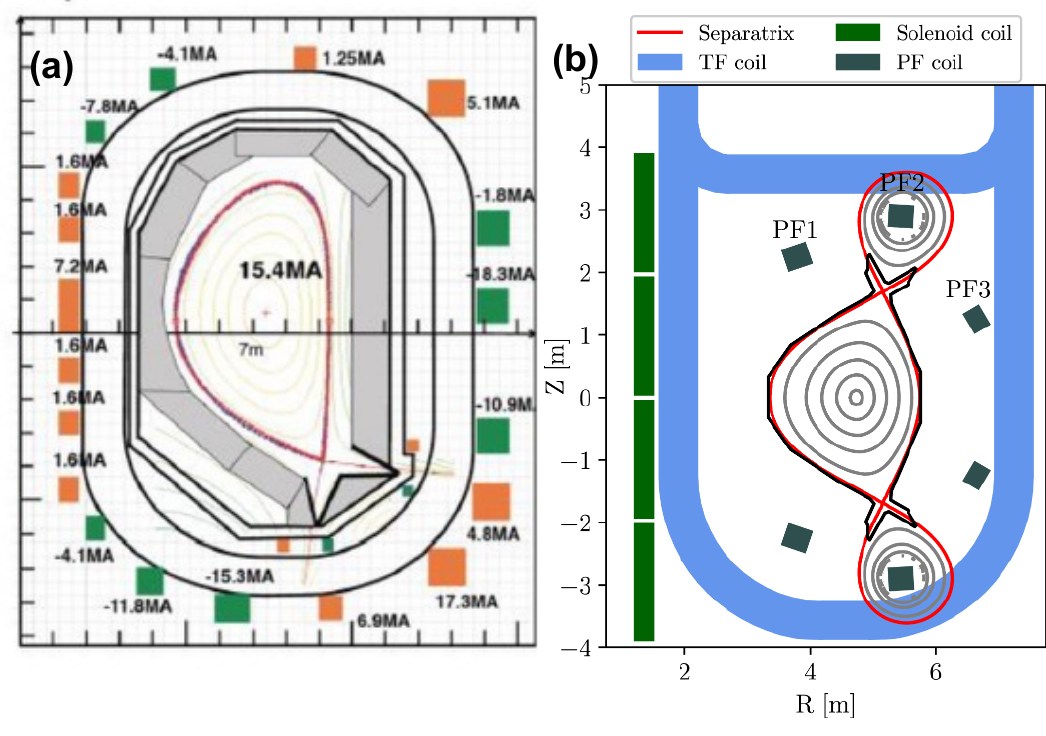}
\caption{\label{fig:figure7} NT FPP designs of a large major radius (a), reproduced from \cite{Kikuchi_2019}, and the MANTA high-field (b),  reproduced from \cite{manta}.}
\end{figure*}
The momentum for the NT reactor path is reflected in several proposed upcoming international projects and private sector initiatives. The planned DTT device in Italy is currently investigating the NT option \cite{Mariani_2024}, supported by experimental results from both TCV\cite{Balestri_2024ppcf} and AUG \cite{Aucone_2024} along with initial gyrokinetic modeling\cite{Mariani_20242}. Also, the possibility of NT shapes in SPARC has recently been assessed \cite{DEBOUCAUD2024114401,Yuksek_2026_SPARC}. NT was first considered as a viable reactor scenario in the 2010s, focusing on the "power-handling first" philosophy characterized by a large major radius and the potential for significant flux expansion \cite{Kikuchi_2019, Medvedev_2015}. An example of such an approach is shown in Fig. 28(a). 

Recently, the focus has shifted toward more compact, high-field configurations, as seen in several student-led NT FPP design studies that make use of high magnetic fields. MANTA (Modular Adjustable Negative Triangularity ARC-class)\cite{manta,Wilson_2025,Miller_2024} is a pulsed, radiative, high-field ELM-free tokamak with $\delta\sim-0.5$; the poloidal cross-section of MANTA is shown in Fig. 28(b). CENTAUR (Compact Experimental Negative TriAngUlarity Reactor) was designed as a feasible, high-field energy breakeven NT tokamak\cite{APS_DPP2025_PP13_139}. There has also been a recent study investigating the possibility of a low power NT FPP\cite{Balestri_2025}. Additionally, there is significant interest from private industry, with companies such as Startorus Fusion \cite{Tan2024NTST} and Firefly Fusion proposing new NT-based machines.

\section{\label{sec:con}Conclusion}

NT has emerged as a promising reactor scenario that addresses the fundamental tension between high core performance and manageable exhaust. While the standard H-mode regime is typically planned for fusion reactors due to its high confinement, it exacerbates exhaust challenges through transient ELM events and suffers from limited reproducibility and accessibility. In contrast, recent experimental progress on devices like TCV and \mbox{DIII-D} has demonstrated that NT provides simultaneous access to high-performance, ELM-free operation in reproducible plasmas. By passively avoiding ELMs, eliminating the L-H power threshold, and utilizing a larger divertor wetted area on the outboard side, the NT path significantly alleviates the power exhaust and impurity flushing constraints that challenge the conventional positive triangularity approach.

The growth of NT experiments worldwide, coupled with the emergence of private industrial efforts and student-led FPP design studies such as MANTA and CENTAUR, underscores the increasing momentum behind this scenario. Although outstanding questions remain, the international research community is rapidly expanding its capabilities to address these gaps. Currently, \mbox{DIII-D} remains the only facility capable of exploring the NT regime at high power in the near-term, and a proposed upgrade to its divertor geometry will be essential for validating the NT path for a pilot plant. Ultimately, the unique physics and engineering advantages of negative triangularity provide a robust and simplified foundation for the development of an economically viable fusion power plant.
\begin{acknowledgments}
This material is based upon work supported by the U.S. Department of Energy, Office of Science, Office of Fusion Energy Sciences, using the DIII-D National Fusion Facility, a DOE Office of Science user facility, under awards DE-FC02-04ER54698, DE-FG02-04ER54761, DE-SC0022270, DE-FG02-97ER54415, DE-AC52-07NA27344, and DE-SC0014264. This work was supported in part by General Atomics corporate funding. This work has been carried out within the framework of the EUROfusion Consortium, funded by the European Union via the Euratom Research and Training Programme (Grant Agreement No 101052200 — EUROfusion) and funded by the Swiss State Secretariat for Education, Research and Innovation (SERI). Views and opinions expressed are however those of the author(s) only and do not necessarily reflect those of the European Union or the European Commission. Neither the European Union nor the European Commission can be held responsible for them.
\end{acknowledgments}

\vspace{0.2in}
{\linespread{1.}\small
\textbf{Disclaimer-}  This report was prepared as an account of work sponsored by an agency of the United States Government.  Neither the United States Government nor any agency thereof, nor any of their employees, makes any warranty, express or implied, or assumes any legal liability or responsibility for the accuracy, completeness, or usefulness of any information, apparatus, product, or process disclosed, or represents that its use would not infringe privately owned rights.  Reference herein to any specific commercial product, process, or service by trade name, trademark, manufacturer, or otherwise, does not necessarily constitute or imply its endorsement, recommendation, or favoring by the United States Government or any agency thereof.  The views and opinions of authors expressed herein do not necessarily state or reflect those of the United States Government or any agency thereof.
\section*{Data Availability Statement}

This is the data availability statement required by policies of the \mbox{DIII-D} National Fusion Facility. The data that support the findings of this study are available upon reasonable request from the authors.

\section*{References}
\bibliography{negd}

\end{document}